\documentclass[11pt,a4paper]{article}
\usepackage{jheppub}
\usepackage{graphicx}
\usepackage{amssymb}
\usepackage{amsmath}
\usepackage{lineno}
\usepackage{placeins}
\usepackage{xcolor}
\usepackage{xspace}
\usepackage{hyperref}
\usepackage{cleveref}
\usepackage{booktabs}
\usepackage[scientific-notation=true]{siunitx}

\usepackage{booktabs}

\newcommand{\todo}[1]{}
\renewcommand{\todo}[1]{{\color{red} TODO: {#1}}}

\newcommand*{\Rinv}{\ensuremath{R_{\textrm{inv}}}}

\newcommand*{\minphi}{\ensuremath{\phi_{\textrm{MET},\textrm{closest-jet}}}}

\def \kt {\ensuremath{k_{t}}\xspace}
\def \antikt {anti-\kt}

\def \pT {\ensuremath{p_{\mathrm{T}}}\xspace}
\def \met {\ensuremath{p_{\text{T}}^{\text{miss}}}\xspace}
\def \metval {\ensuremath{E_{\text{T}}^{\text{miss}}}\xspace}

\graphicspath{{Figures/}}

\title{Semi-visible jets from Sneaky dark matter}

\author[a]{Deepak~Kar\footnote{In memoriam.}} 
\author[b]{\\Christiane Scherb \footnote{cscherb@lbl.gov}}
\author[c]{\\Pedro Schwaller \footnote{pedro.schwaller@uni-mainz.de}}
\author[d]{\\Sukanya~Sinha \footnote{sukanya.sinha@cern.ch}}

\affiliation[a]{School of Physics, 
University of Witwatersrand, Johannesburg, South Africa, and \\ 
Royal Society Wolfson Visiting Fellow at the University of Glasgow, United Kingdom}

\affiliation[b]{Leinweber Institute for Theoretical Physics, Department of Physics, University of California, Berkeley, CA, USA, and \\
Theoretical Physics Group, Lawrence Berkeley National Laboratory, Berkeley, CA, USA}

\affiliation[c]{PRISMA$++$ Cluster of Excellence \& Mainz Institute for Theoretical Physics,
Johannes Gutenberg University, Mainz, Germany}

\affiliation[d]{School of Physics and Astronomy, University of Manchester, Manchester, United Kingdom}

\abstract{
In this paper, we compare two different ways of generating semi-visible jet signature
in non-resonant production mode, and flag the potential advantages of the new generation approach we propose. 
In addition, we suggest a more efficient matrix element 
to parton shower matching configuration for this hadronic final state signature. Finally, we perform a reinterpretation
study of the ATLAS search for t-channel semi-visible jet production and obtain collider limits on the sneaky DM scenario.  
}

\makeatletter
\gdef\@fpheader{}
\makeatother

\begin{document}
\maketitle



\section{Introduction}
\label{sec:intro}

Dark sectors containing a new confining gauge force are well motivated from both the theoretical and experimental perspectives \cite{Kopp:2016yji,Essig:2009nc,Bhattacharya:2013kma,Cline:2013zca,Hochberg:2014kqa,Harigaya:2016rwr,Berlin:2018pwi,Beauchesne:2018myj,Beauchesne:2019ato,Bernreuther:2019pfb,Contino:2020god,Chu:2024rrv,Garcia-Cely:2024ivo,Maleknejad:2022gyf,Alexander:2023wgk,Alexander:2024nvi,Manzari:2022iyn,Renner:2018fhh,Carmona:2021seb,Carmona:2022jid,Jubb:2017rhm,Agrawal:2014aoa,Albouy:2022cin,Strassler:2006im,Han:2007ae,Butterworth:2023cgz,Cohen:2020afv,Cohen:2023mya,10.21468/SciPostPhysCore.7.4.071,Batz:2023zef,Beauchesne:2022phk,Strassler:2008fv,Cazzaniga:2022hxl,Bernreuther:2020vhm,Park:2017rfb,Bertone:2018krk}. 
Confining dark sectors with similar properties as standard model QCD are often termed dark QCD.
A dark QCD sector can contain different flavours of dark quarks, that subsequently form bound states called dark hadrons that may be either stable or unstable. 
The stable dark hadrons can be potential dark matter (DM) candidates that will remain invisible to the detector, while the
unstable dark hadrons can decay to standard model (SM) particles. 
This can lead to a plethora of unique collider final state signatures, including semi-visible jets \cite{Cohen:2015toa,Cohen:2017pzm,Beauchesne:2017yhh} and emerging jets \cite{Schwaller:2015gea,Mies:2020mzw,Linthorne:2021oiz,Carrasco:2023loy}. 
Here, we will concentrate on the 
semi-visible jets (SVJs) signature, 
where a subset of the dark hadrons remain stable and contribute to missing transverse momentum, while the rest decay promptly to SM particles. 

SVJs can be generated via either resonant or non-resonant mediators. 
This paper will focus on the non-resonant ($t$-channel) production of semi-visible jets and will compare two different Unified Feynrules Output (UFO) models that can currently be used for the matrix element level generation. 
The comparison will be followed by a discussion of the features 
of using each of those UFOs, and finally we will conclude with 
recommendations on the way forward for generating SVJs via a non-resonant mediator.

\begin{figure}[ht]
  \centering
  \begin{minipage}[b]{0.33\textwidth}
    \centering
    \raisebox{0.9cm}{\includegraphics[width=\linewidth]{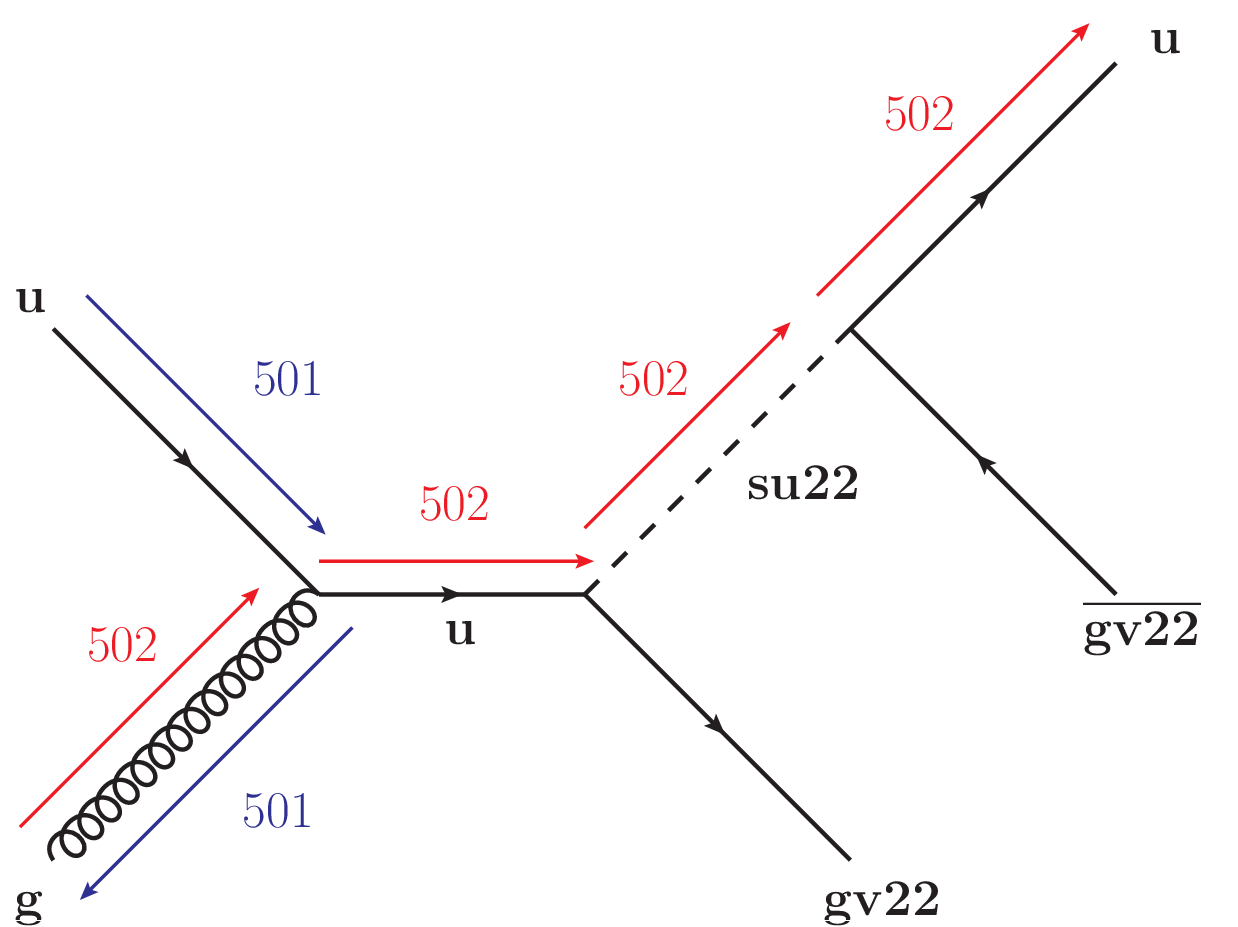}}
  \end{minipage}\hspace{0.07\textwidth}%
  \begin{minipage}[b]{0.33\textwidth}
    \centering
    \includegraphics[width=\linewidth]{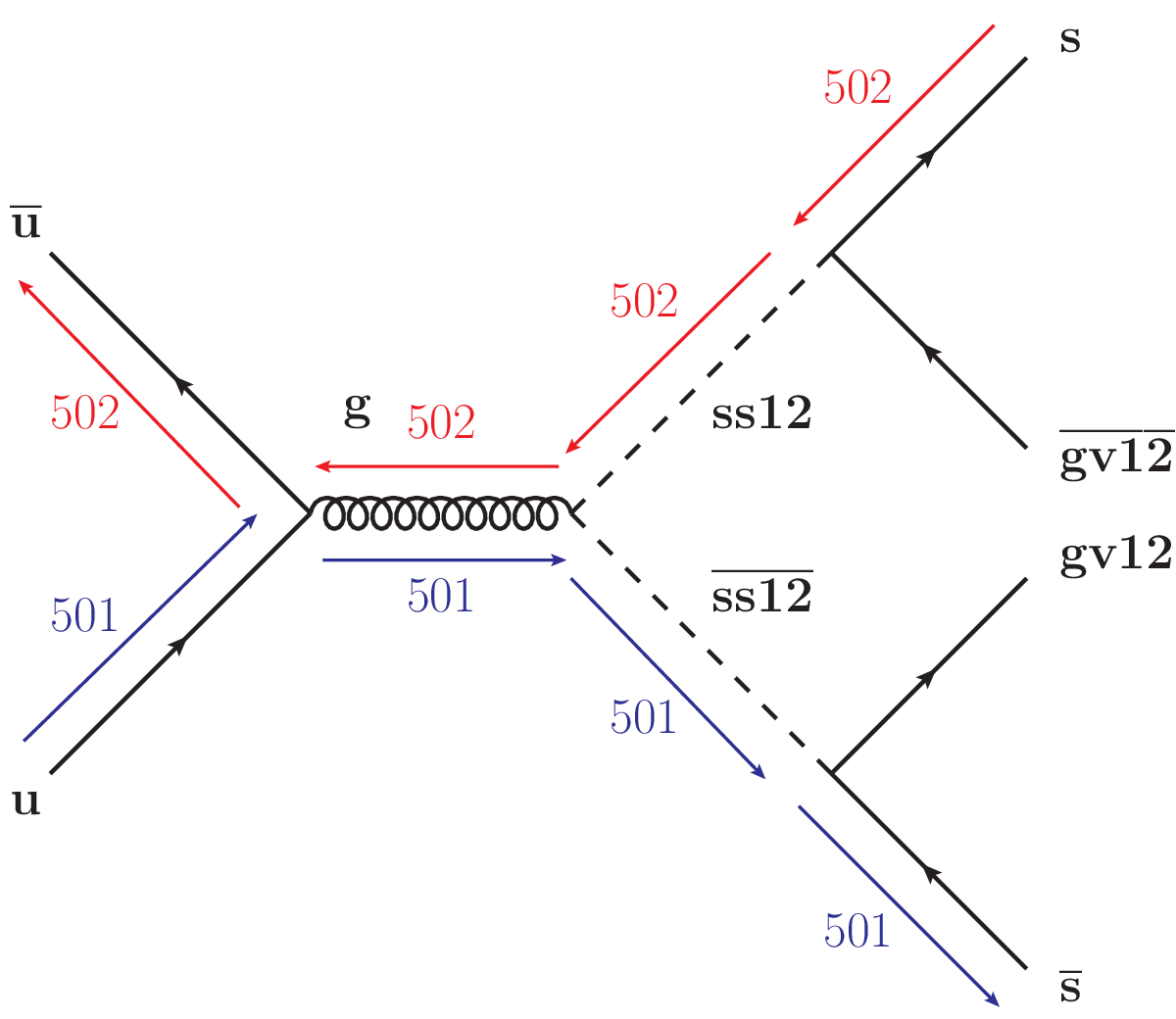}
  \end{minipage}

  \caption{Feynman diagrams for generating $t$-channel production of two dark quarks, with one (left) or (two) extra partons, with SM colour flow lines indicated. The $su22$ or the $ss12$ are the mediators, and $gv12$ or $gv22$ are the dark quarks.}
  \label{fig:FDcohen}
\end{figure}

The ATLAS Full Run-2 non-resonant search~\cite{ATLAS:2023swa} 
used the UFO model listed in Ref.~\cite{dmsimp}. Throughout this paper we will refer to it as Model1. 
The parton level events are generated 
in Madgraph~\cite{Alwall:2014hca, Frederix:2012ps} and the resulting dark quarks subsequently undergo dark shower and hadronisation using the Pythia\,8~\cite{Sjostrand:2014zea} 
Hidden Valley (HV)~\cite{Carloni:2011kk} parton shower (PS) module. 
The dark hadronisation gives rise to  flavour-diagonal and 
off-diagonal $\pi_{\mathrm{d}}$ and $\rho_{\mathrm{d}}$ mesons, with spin~0 and~1 respectively.
However, this approach has certain limitations. 
The UFO model listed in Ref.~\cite{dmsimp} implements a 
generic $t$-channel dark matter (DM) model featuring multiple
bi-fundamental mediators which decay into quark - dark quark pairs. 
However, here the bi-fundamental 
mediators do not contain any true HV colours, and the resulting DM particles do not have any inherent HV quantum numbers as well. 
The particle IDs of the DM particles need to be changed to match the HV dark quarks. 
Typically, the signal process for non-resonant SVJ searches is generated as two dark quarks forming SVJ with up to two extra partons at parton level. 
The MLM~\cite{Mangano:2006rw} jet matching scheme, 
with the matching parameter set to 100~GeV was used.
While this method worked till Pythia\,8306, 
the later versions of Pythia\,8 incorporated a stricter HV colour counting, 
which resulted in generation errors.
Fig.~\ref{fig:FDcohen} shows example processes with one and two extra partons. 
The QCD colour flows are indicated by red and blue arrows as per Pythia\,8 event record. 
The $su22$ or the $ss12$ are the mediators, and $gv12$ or $gv22$ are the dark quarks accoring to the naming scheme of Model1. 
Since these mediators do not have any HV colour, 
in order to maintain the HV colour flow (not shown), 
Pythia\,8 has to be explicitly instructed to consider them as point interactions. 
However, that complicates the QCD colour counting 
and the event generation does not succeed.


\begin{figure}[ht]
  \centering
  \includegraphics[width=0.33\textwidth]{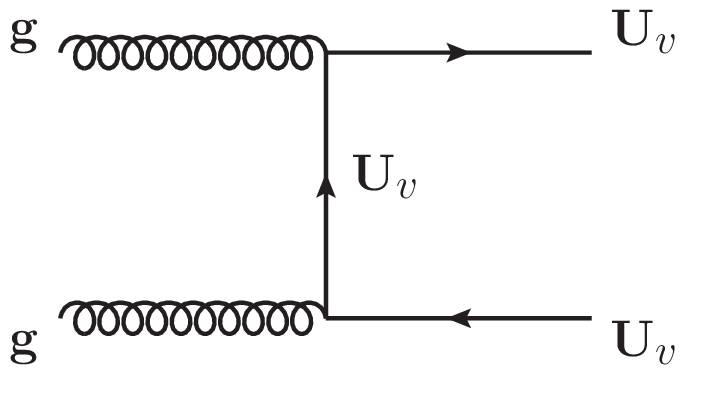}
  \caption{Feynman diagram for generating a pair of 
  dark quarks, $D_v$ in $t$-channel from two incoming gluons.}
  \label{fig:FDnaive}
\end{figure}

An alternate way would be to generate the $t$-channel 
process naively in Pythia\,8 HV, 
following the approach adapted in~\cite{Bhardwaj:2024djv} by using \texttt{HiddenValley:gg2UvUvbar} 
where two incoming gluons produce a pair of bi-fundamental mediators, e.g. $U_v$ as in Fig.~\ref{fig:FDnaive}. 
The HV particle $U_v$ is a partner to the $u$ quark, and is a part of 
the set of 12 particles that mirrors the Standard Model flavour structure. These 12 HV particles 
are charged under both the SM and the HV symmetry groups, with quantum numbers matching the $su22$ and related states of Model1.
The $U_v$ can decay to a $u$ quark and a HV quark, 
which shower and hadronise, and can result in dark hadrons forming a SVJs. Setting the mass of $U_v$ to the TeV scale it can play the role of the mediator for $t$-channel event generation. However in this setup it is impossible to simulate the production of additional hard jets, and to incorporate additional production channels such as the one shown in Fig.~\ref{fig:FDcohen} (left). Once initial and final state radiation is included in the simulation, it furthermore becomes necessary to be able to perform matching between hard additional jets and soft radiation, which is not possible with the hard coded Pythia processes. For this, a benchmark model with a working UFO file is required, such that all tree level diagrams up to a given jet multiplicity can be calculated with a matrix element calculator such as MadGraph, and to match with ISR and FSR radiation during event generation.

Incidentally, a suitable UFO benchmark model was implemented for the sneaky DM model~\cite{dmsneaky}, where a single bi-fundamental mediator connects quarks and dark quarks.\footnote{An additional approach to generate SVJs by using dark glueballs has 
been proposed in~\cite{Batz:2023zef}, 
but it is still under investigation.} In the following we will refer to this as Model2. The goal of the present paper is to establish Model2 as benchmark model for future semi-visible jets searches by comparing event generation in both models and recasting the existing searches onto Model2. 
In section~\ref{sec:models} we introduce the underlying theories of the two UFO models, followed by a systematic comparison of the event generation in section~\ref{sec:comparsion}. In section~\ref{sec:matching} we recommend a more efficient matching setup for the generation of SVJ jets, before showing a re-interpretation of the ATLAS run-2 SVJ results~\cite{ATLAS:2023swa} using both Model1 and Model2 in section~\ref{sec:reint}. We conclude in section~\ref{sec:summ}.

\section{Model comparison}
\label{sec:models}
The UFO model files~\cite{dmsimp}, Model1, are based on the following interaction Lagrangian~\cite{Cohen:2017pzm}:
\begin{align}
    {\cal L}_1 \supset \sum_{ijab} \lambda_{ijab} \bar{Q}_a \Phi^*_{bi} q_{{\mathrm R}j}\,. \label{eqn:lagrangian1}
\end{align}
Here $q_{{\mathrm R}i}$ are right-handed SM quarks $i\in\{u,d,s,c,b,t\}$, $\bar{Q}_a$ are dark quarks, i.e. Dirac fermions which are singlets under the SM interactions but transform in the fundamental representation of the dark gauge group, and $a\in\{1,2\}$ labels the two dark flavours. The bi-fundamental scalars $\Phi_{bi}$ are the mediators, with appropriate charges such that the interaction term~\cref{eqn:lagrangian1} is gauge invariant. There are in total \textbf{12 complex scalars} which are assumed to have a common mass $M_\Phi$, and the coupling matrix takes the form $\lambda_{ijab} = \lambda \delta_{ij}\delta_{ab}$.  Note that in the UFO model the real and complex parts of the mediator are implemented separately, so that there are effectively 24 mediators. Similarly, the dark quark parts coupling to the real and complex part of the mediator are implemented separately leading to effectively four dark quarks. In the UFO model the bi-fundamental mediators are named $s\langle qij\rangle$, where q is the SM quark to which a given mediator couples, and $ij$ the indices of the respecitve dark quark, see Fig.~\ref{fig:FDcohen} for an example.

Recently a composite DM model, based on a $SU(N_d)$ dark sector with 
dark quarks and a heavy bi-fundamental $t$-channel mediator consistent with Pythia\,8 HV setup has been proposed~\cite{Carmona:2024tkg}, with the UFO files available in Ref~\cite{dmsneaky}. 
The sneaky DM model introduces dark quarks $Q_\alpha$ and a bi-fundamental scalar mediator $X$ with interactions 
\begin{align}
    {\cal L}_2 \supset -\sum_{\alpha i} \kappa_{\alpha i} \bar{q}_{{\mathrm{R}i}} XQ_\alpha \,. 
\end{align}
Here, $\alpha \in\{1,\dots,n_f\}$ labels the dark quark flavours, and $i\in\{1,2,3\}$ labels either SM up-type or down-type quarks, depending on the quantum numbers of the \textbf{single} complex scalar mediator $X$. Since there is only one mediator, the hypercharge of the mediator determines, if it couples to up-type ($Y_X = -\frac{2}{3}$) or down-type quarks ($Y_X= \frac{1}{3}$). In this model, for $n_f\ge4$ some dark pions carry a dark flavour charge making them stable and the DM candidate. Therefore, for the phenomenological studies the coupling matrix can be taken as $\kappa_{\alpha i} = \kappa$ for $i = 1,2,3$ and $\alpha = 1,2,3$, while $\kappa_{\alpha i} = 0$ for $\alpha > 3$. A different choice of coupling structure will lead to different factors between the two models in section~\ref{sec:comparsion}. 
In the following we will refer to this model as Model2.

While the two interaction terms have very similar structure, it should be clear now that in practice the two models are very different. Most importantly, the dependence of the cross section on the couplings $\kappa$ and $\lambda$, respectively, is non-trivial and can not be absorbed in a trivial rescaling everywhere, since some contributions also scale with the number of mediators, while others are determined by the strong coupling. The goal of the next section is to perform a more detailed comparison and to understand how existing limits can be reinterpreted consistently in Model2. Unlike Model1, the single mediator in Model2 allows flavour changing  interactions. Therefore, in a realistic model setup bounds from flavour have to be considered for Model2-like scenarios \cite{Carmona:2021seb,Carmona:2024tkg, Renner:2018fhh}.

\section{Comparison between the two models}
\label{sec:comparsion}

\subsection{Generation setup}
\label{sec:generation}

We follow the analysis in~\cite{ATLAS:2023swa} and produce pairs of dark quarks with no, one or two additional SM jets. For Model2 $Y=-\frac{2}{3}$ is chosen for the hypercharge of the mediator, coupling the dark sector to up-type quarks. Both stable and unstable dark hadrons are produced in the dark shower after hadronization. The unstable states further decay either to stable dark hadrons  or to SM quarks, resulting in semi-visible jets, which contain SM hadrons as well as some amount of missing energy. 

In the context of this study, we assume the amount of missing energy depends on an input ratio termed $\Rinv$, which is defined as 
$$\Rinv = \langle\# \text{ stable dark hadrons}\rangle/\langle\# \text{ dark hadrons}\rangle\,.$$

To be precise, the number in the denominator should only contain dark hadrons which are either stable or decay dominantly to SM particles, such that this quantity properly counts the average missing energy per SVJ. In a given model $\Rinv$ is in principle calculable, and in the simplest cases can only assume discrete values. However it is relatively easy to introduce additional interactions which change $\Rinv$ without affecting other aspects of the model. To not artificially exclude such scenarios from searches, we therefore treat $\Rinv$ as a phenomenological parameter that can assume values between 0 and 1.

As heavier dark hadrons promptly decay to lighter ones, for the collider final states we focus only on dark pions $\pi_D$ and dark rho mesons $\rho_D$.
Specifically, in the current work, we assume that flavour-diagonal $\pi_{\mathrm{d}}$ and $\rho_{\mathrm{d}}$ 
mesons are unstable, while off-diagonal $\pi_{\mathrm{d}}$ and $\rho_{\mathrm{d}}$ are stable. The unstable diagonal dark pions and rho mesons decay promptly to stable, off-diagonal $\pi_D$ and $\rho_D$ pairs respectively, and to SM quarks. 

For event generation at parton level we use MadGraph5$\_$aMC@NLO version 3.5.15, while shower and hadronisation is carried out with the Hidden Valley module of Pythia8. 
For Model1 we use Pythia\,8.306 since later versions are not able to run the generated Les Houches Event file, and for Model2 we use Pythia\,8.313 as earlier versions do not have the three flavours of dark quarks used in this model. In order to be consistent, the following HV parameters were set identically in Pythia as a starting point:

\begin{itemize}
\item the dark QCD confinement scale value ($\Lambda_{D}$), set at 5 GeV.
\item the number of dark colours ($N_C$), set at 3.
\item the number of dark quark flavours ($N_F$), set at 2.
\item the lowest allowed \pT of emission, set at 5.5 GeV.
\item the fraction of HV-mesons that are assigned spin 1 (vector), 
with the remainder spin 0 (pseudoscalar), set at 0.58.
\item the masses of the dark quarks ($M_{q_{D}}$), set at 10 GeV, the same dark quark mass is set in Madgraph.
\item the masses of the diagonal, unstable dark hadrons ($M_{\rho_{D}}$, $M_{\pi_{D}}$), set at respectively 10 and 16 GeV
for unstable mesons, and 4.99 and 7.99 GeV for off-diagonal, stable ones.
\end{itemize}

The masses of the dark hadrons were chosen  
so that the above described decays of diagonal, unstable dark hadrons to off-diagonal, stable dark hadrons are kinematically possible. It was, however, observed that these and other detailed choices made for 
the dark sector parameters did not influence the final state topology. The parameter choices used here, are not the same as were proposed in~\cite{Albouy:2022cin}, rather the parameter values are chosen to correspond to the ATLAS analysis in \cite{ATLAS:2023swa}, though the generation settings are slightly different. 

Based on this setup the process $p p\to Q_D \bar{Q}_D$ is part of the production processes for SVJs. As this process is mediated by the t-channel exchange of a heavy mediator, its cross section develops a collinear mass singularity as $m_{Q_D}\to 0$. This is a general feature of t-channel exchange with a chiral coupling to a light final-state fermion and leads to unstable phase-space integration in Madgraph. To get well-defined cross sections we regulate this divergence with a $p_T$ cut on the dark quarks. Throughout this work we impose $p_T^{Q_D} > 100$~GeV, though we also give some examples with $p_T^{Q_D} > 200$~GeV.


\subsection{Comparisons at Parton Level}
\label{sec:partoncomp}

As described in section~\ref{sec:models} the two models considered differ in the number of mediators,  dark quark flavors and, consequently, the coupling structure. To enable a direct comparison between them, we need to find parameters for which the generated signals are similar. In this section, we systematically compare the production of SVJs in the two models, starting with simple $2\to2$ processes on parton level, and eventually moving to the full process considered in the ATLAS search in~\cite{ATLAS:2023swa}.

\paragraph{$\boldsymbol{pp\to Q_D Q_D}$}
We first consider only $2 \rightarrow 2$ processes at parton level, specifically $p p \to Q_D \bar{Q}_D$ with a mediator exchanged in the t-channel.  Here, $Q_D$ corresponds to the two dark quarks $Q_a$ (a=1,2) of Model1, or the three dark quark flavors $Q_\alpha$ ($\alpha=1,2,3$) in Model2 that couple to SM quarks. 
In Model1 there are 16 processes contributing to $p p \to Q_D\bar{Q}_D$, originating from diagonal couplings ($\lambda_{ijab} = \lambda \delta_{ij}\delta_{ab}$) of SM quarks $q = u,d,c,s$ to the two dark quarks via the respective complex mediators. 
For Model2, on the other hand, there are $n_q^2 \times n_{Q_D}^2 =36$ processes contributing, corresponding to all combinations of two initial light  SM quarks, $u,\,c$  ($d,\,s$) for couplings to up-type (down-type) quarks, coupling to all possible combinations of two dark quarks in the final state ($\kappa_{\alpha i} = \kappa$).
As we assume degenerate dark quark masses and mediator masses, respectively, as well as universal couplings, all sub-processes have the same matrix element. 
Thus, for $\lambda=\kappa = const.$, we find a ratio of $9/4$ between the number of sub-processes contributing to $p p \to Q_D\bar{Q}_D$ in Model2 and Model1. Because cross sections of all contributing processes scale as $coupling^4$, choosing the coupling $\kappa$ of Model2 to be $\kappa=\lambda\times\left(\frac{4}{9}\right)^{1/4}=\lambda\times\sqrt{\frac{2}{3}}$ is expected to give the same cross sections as Model1 up to parton distribution functions (PDFs) corrections due to the different PDFs for the initial state quarks. 

\begin{figure}[ht]
    \centering
    \includegraphics[width=0.67\linewidth]{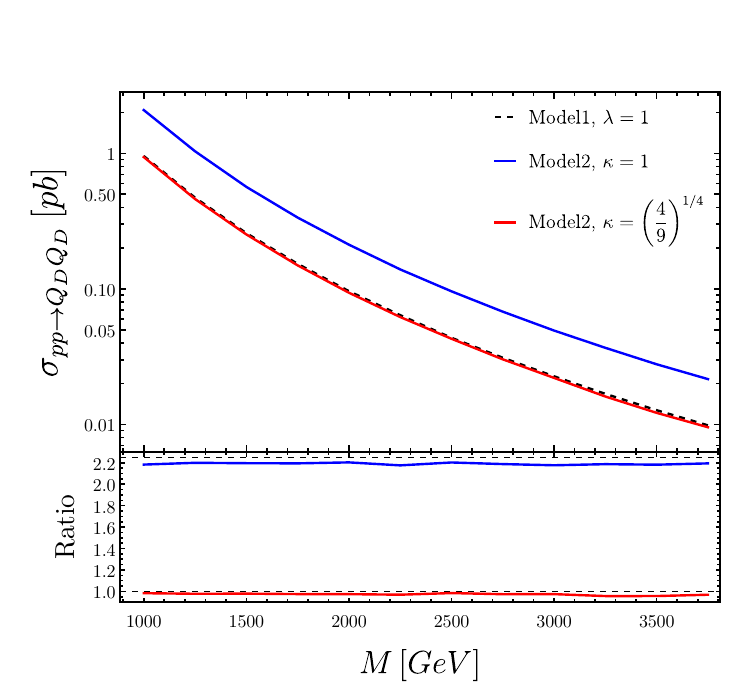}
    \caption{$p p \to Q_D Q_D$ cross section as a function of the mediator mass $M$ at $\sqrt{s}=13$~TeV for $\lambda =1$ for Model1 (black, dashed) and for $\kappa=\lambda$ (blue) and $\kappa = \lambda\times \left(\frac{4}{9}\right)^{1/4} =0.817$ (red) for Model2. A \pT $>$ 100~GeV cut has been applied to all dark quarks. The ratio of the Model2 cross sections to the Model1 cross section is shown in the lower panel. Dashed lines in the lower panel correspond to the expected ratios 1 and 9/4.}
    \label{fig:xsecs}
\end{figure}

We show the cross sections calculated using Madgraph for $p p \to Q_D \bar{Q}_D$ at $\sqrt{s}= 13$~TeV and the minimum \pT requirement for dark quarks set at \pT$>100$~GeV in Fig.~\ref{fig:xsecs}. The cross section for Model1 with $\lambda =1$ is displayed as the black, dashed line, the one for Model2 with couplings to up-type quarks with $\kappa=\lambda = 1$ and $\kappa = \lambda\times \left(\frac{4}{9}\right)^{1/4} =0.817$ in blue and red, respectively. In the lower panel the ratios of the cross sections are shown together with the expected ratios 9/4 for $\kappa = 1$ and 1 for $\kappa=0.817$ as dashed black lines. We find, that the cross sections exhibit the expected behavior. The Model2 cross section with $\kappa = 1$ is larger than the one for Model1 by roughly the expected factor of $9/4$, while they agree for $\kappa =0.817$ within a few percent. For both choices of $\kappa$ the ratio of the cross sections calculated in Madgraph is slightly below the expected one, which is likely due to the different PDFs contributing to the process. More precisely, Model1 benefits from the up and down quark PDF (and sub-leading the strange and charm PDFs), while Model2 only picks up the up and charm PDFs.   

\begin{figure}[ht]
\centering
\includegraphics[width=0.45\textwidth]{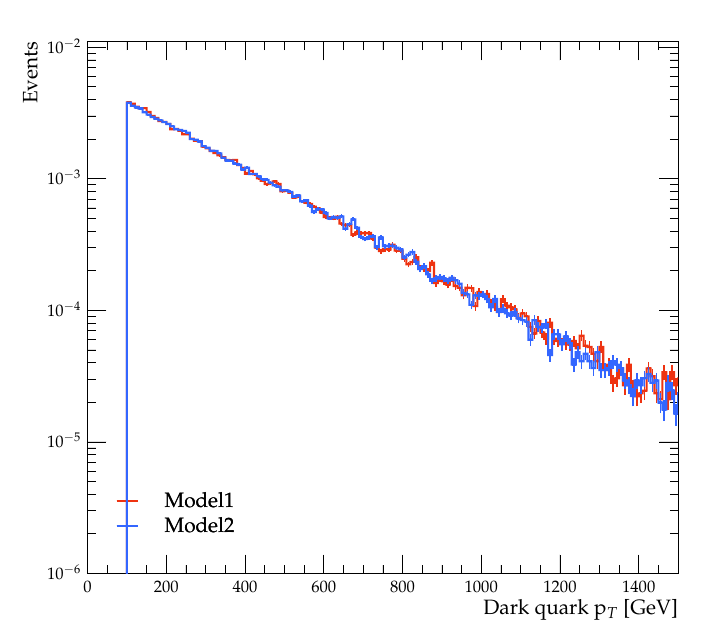}
\includegraphics[width=0.45\textwidth]{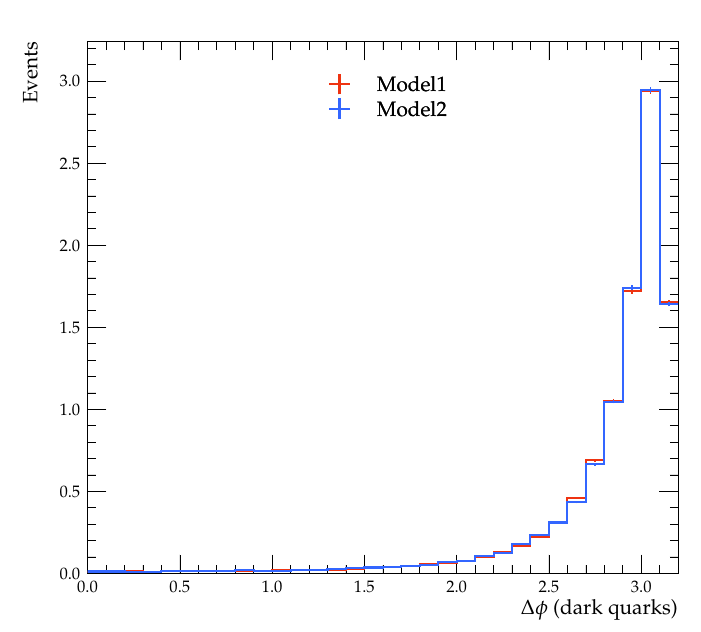} 
\caption{Parton level Model1 against Model2 comparisons for $p p  \rightarrow Q_D \bar{Q}_D$ with a mediator mass  of $M=3000$~GeV, and \Rinv\ = 0.5, for dark quarks \pT and azimuthal angular difference.}
\label{fig:comp-parton}
\end{figure}

Next, we compare some kinematic variables for the same process. As before we require \pT$>100$~GeV for the dark quarks, and apply no other parton level cuts. The final state partons were identified using their PID. For a representative mediator mass of $M=3000$~GeV with  $\Rinv =0.5$ the parton level dark quark \pT and azimuthal angular difference is shown in Fig.~\ref{fig:comp-parton}.  We use $\lambda=\kappa=1$ and normalize from here on all distributions to area. As expected, almost a perfect agreement between Model1 and Model2 is observed. 

\paragraph{$\boldsymbol{pp\to X X^\dagger}$}
Before we move to a comparison of the full process a comment on the second possible $2\to2$ process, $p p \to X X^\dagger/\Phi_{bj}\Phi_{ai}^\dagger$, is in order. Mediator pair production can happen in two ways: through QCD interactions, when a mediator pair is produced from gluons, and via a t-channel exchange of a dark quark \cite{Mies:2020mzw,Carmona:2024tkg}. 
For $\kappa=\lambda\lesssim0.1$ the QCD contributions to $p p \to X X^\dagger/\Phi_{bj}\Phi_{ai}^\dagger$ dominate. Since there are no dark sector couplings in the QCD processes, a simple rescaling of the couplings to match the cross sections of the two models is not possible. Nonetheless, we can understand the difference in cross sections for small couplings by comparing the number of production processes: In Model1 there are 192 pure QCD processes and 64 t-channel processes, while in Model2 there are only 8 pure QCD processes and 12 t-channel processes. 
We recover the expected ratio in cross section of $8/192=1/24$ between the Model2 and Model1 cross sections for $\lambda=\kappa=0.1$. 

For $\lambda=\kappa\sim\mathcal{O}(1)$ t-channel processes contributions become relevant and this scaling is no longer applicable. In the ATLAS SVJ search \cite{ATLAS:2023swa} the processes $p p \to Q_d \bar{Q}_D, p p \to Q_d \bar{Q}_D j,  p p \to Q_d \bar{Q}_D j j$ are considered. The latter includes $p p \to X X^\dagger/\Phi_{bj}\Phi_{ai}^\dagger$ among other processes making a simple scaling of the cross sections with the coupling or the number of processes non-trivial. We will discuss the different contributions in detail in the next section. 

For emerging jets searches, on the other hand, often only production via $p p \to X X^\dagger$ or $p p \to\Phi_{bj}\Phi_{ai}^\dagger$ is considered \cite{ATLAS:2025lfx, CMS:2024gxp}. Then, for small couplings the naive $1/24$ scaling can be used for reinterpretation, while full cross section calculations are needed for larger couplings.

\paragraph{\textbf{Full process}}

Finally, we compare the cross sections for the full process of the SVJ search ($p p \to Q_D \bar{Q}_D, p p \to Q_D \bar{Q}_D j,  p p \to Q_D \bar{Q}_D j j$) for both models for different mediator masses and couplings in Tab.~\ref{tab:xsec}. All cross sections were determined with a parton level cut $p_T^{min} > 100~(200)$~GeV for the dark quarks. 
In the upper part of the table the mediator couplings are constant $\lambda=1$ and $\kappa=0.817$, while in the lower part the mediator mass is kept fixed at $m_X=3000$~GeV and the couplings are varied. 

We find that for $\lambda = 1,\, \kappa = 0.871$ for mediator masses $M\geq2$~TeV the cross sections agree within a few percents. 
Only for $M=1$~TeV there is a larger disagreement between the two cross sections of $\sim 30\%$. This effect can be attributed to the relative contribution of the pure QCD mediated pair production cross section, that contributes to the $p p \to Q_D\bar{Q}_Djj$ process, to the total cross section. 
Above the kinematic threshold $\sqrt{\hat{s}}\geq 2M$, where $\sqrt{\hat{s}}$ is the parton center-of-mass energy, the mediator pair production cross section is described by a Breit-Wigner distribution
, peaking close to threshold and then falling off with $(M-\sqrt{\hat{s}})^2$. With increasing mediator masses the kinematic threshold increases and at the same time the pair production cross section decreases. For processes with the mediator in the t-channel, on the other hand, the cross section increases logarithmically as a function of the initial parton center-of-mass energy unless regulated by an $|\eta_{Q_\alpha}|$ cut. Even with the regularization the cross sections fall slower with $\sqrt{\hat{s}}$ than the QCD mediator pair production and the kinematic threshold is independent of the mediator mass. Combined with the fact that the gluon PDF decreases for larger energy fractions and is overtaken by the up and down quark PDFs (relevant for the t-channel processes), the pure QCD mediator pair production has a larger relative contribution to the total cross section at small mediator masses, leading to a larger difference in the cross sections of the two models for $M\lesssim2$~TeV. 

We observe similar behaviour for decreasing $\lambda,\, \kappa$ in the lower part of Tab.~\ref{tab:xsec}: At $\lambda=0.5$, $\kappa =0.408$ we find good agreement between the cross sections of the two models within a few percent, while they diverge for smaller couplings until around $\lambda\sim0.075$, $\kappa\sim0.061$. Noteworthy, for smaller couplings ($\lambda\sim0.05$, $\kappa\sim0.041$)  the cross section ratio increases slightly. This behavior again stems from the different relative contribution of processes to the full cross section. For larger couplings ($\gtrsim0.5$ for our benchmark point $m_X = 3$~TeV) the three processes $pp\to Q_D Q_D$, $pp\to Q_D Q_D j$ and $p p \to Q_D Q_D j j$ contribute nearly equally to the total cross section and are dominated by parton level processes that scale as $coupling^4$. Therefore, the naive coupling scaling leads to agreement of the cross section between the models. For smaller couplings $p p \to Q_D Q_D j j$ becomes the dominant contribution to the total cross section. In addition to the pure QCD process ($g g \to X X^\dagger \to Q_D Q_D j j$), processes with one gluon and one quark, and two quarks in the initial state also contribute as shown in Fig.~\ref{fig:res_enhancement}. Consequently, the full cross sections contains processes $\propto coupling^4$, $\propto coupling^2$ and independent of the coupling. The process $pp\to Q_D Q_D j$ has contributions  $\propto coupling^4$ and $\propto coupling^2$, but the contribution $\propto coupling^4$ dominates down to smaller couplings than for $p p \to Q_D Q_D j j$, reflecting its relatively larger weight and making $p p \to Q_D Q_D j$ sub-leading. Due to the differences in the two models the relative contributions of processes $\propto coupling^4$, $\propto coupling^2$ and independent of the coupling to $p p \to Q_D Q_D j j$ differ, so that the cross section for Model1 changes from a $\propto coupling^4$ slope to a $\propto couping^2$ slope at larger couplings than the Model2 cross section. This leads to the cross section ratio first decreasing before increasing again for decreasing couplings. 

Finally, it should be noted, that for couplings below \textit{a few times} $10^{-2}$ the decay width  of the mediator(s) becomes smaller than $\Lambda_{QCD}$ and the mediator width can no longer be reliable calculated from its tree-level decays. 
Therefore, we do not show smaller coupling scenarios in Tab.~\ref{tab:xsec}. 
Madgraph avoids the case of $\Gamma\lesssim\Lambda_{QCD}$ by setting the width(s) to zero, effectively making the mediators stable.

\begin{table}[t]
{\footnotesize 
\centering
\begin{tabular}{cccrlrlc}
\toprule
$M$ [GeV] & $\lambda$ & $\kappa$ & \multicolumn{2}{c}{$\sigma$ [pb] Model1} & \multicolumn{2}{c}{$\sigma$ [pb] Model2} & ratio M2/M1 \\
\midrule
1000 & 1& 0.817& 4.03 & (2.65)& 2.90 & (1.83)& 0.719 (0.691)\\
2000 & 1& 0.817& $2.73\times10^{-1}$ & ($2.04\times10^{-1}$) & $2.66\times10^{-1}$ & $1.96\times10^{-1}$ & 0.974 (0.961)\\
3000 & 1& 0.817& $5.75\times10^{-2}$ & ($4.36\times10^{-2}$) & $6.10\times10^{-2}$ & ($4.68\times10^{-2}$) &  1.061 (1.073)\\
4000 & 1& 0.817& $1.92\times10^{-2}$ & ($1.49\times10^{-2}$) & $2.08\times10^{-2}$  & ($1.62\times10^{-2}$) &1.083 (1.087) \\
5000 & 1& 0.817& $8.29\times10^{-3}$ & ($6.51\times10^{-3}$) & $8.87\times10^{-3}$ & ($7.04\times10^{-3}$) & 1.070 (1.081)\\
\midrule
3000 & 0.5& 0.408 & $3.99\times10^{-3}$ & ($3.19\times10^{-3}$) & $3.86\times10^{-3}$  & ($2.97\times10^{-3}$) & 0.967 (0.931) \\
3000 & 0.25& 0.204 & $3.73\times10^{-4}$ & ($3.12\times10^{-4}$) & $2.84\times10^{-4}$  & ($2.25\times10^{-4}$) & 0.761 (0.721) \\
3000 & 0.1& 0.082 & $2.87\times10^{-5}$ & ($2.46\times10^{-5}$) & $1.49\times10^{-5}$ & ($1.28\times10^{-5}$) & 0.519 (0.520) \\
3000 & 0.075& 0.061 & $1.48\times10^{-5}$ & ($1.15\times10^{-5}$) & $6.83\times10^{-6}$ & ($5.95\times10^{-6}$) & 0.462 (0.517) \\
3000 & 0.05& 0.041 & $7.07\times10^{-6}$ & ($6.49\times10^{-6}$) & $4.12\times10^{-6}$ & ($3.64\times10^{-6}$) & 0.583 (0.561) \\
\bottomrule
\end{tabular}
}
\caption{Comparison of cross sections of the full process for SVJ searches for Model1 and Model2 using $p_T^{min} > 100~(200)$~GeV.} 
\label{tab:xsec}
\end{table}

\subsection{Comparisons at Particle Level}
\label{sec:particlecomp}

After finding the coupling scaling that gives comparable cross section for both models, and establishing that at parton level, the basic kinematic distributions agree, a detailed comparison is made at particle level, starting with the same $2 \rightarrow 2$ process. Particle level here means after dark sector shower, hadronization and decays, and SM shower and hadronization.
Again, the minimum \pT requirement for dark quarks was set at 100 GeV. In addition we demand $\pT^{jet} > 25$~GeV and $|\eta^{jets}|< 2.8$ for all jets. We generated events using Madgraph5 with $\sqrt{s}=13$~TeV with showering and hadronization carried out using Pythia8 as described above. 
The same representative value of mediator mass  of $M= 3000$~GeV with  \Rinv\ = 0.5 is compared. 

\paragraph{$\boldsymbol{pp\to Q_D Q_D}$}
In Fig.~\ref{fig:comp-particle2to2} the comparison of six key kinematic observables is shown for the $2\to2$ process $p p\to Q_D\bar{Q}_D$. Namely we show the magnitude of missing transverse momentum $\pT^{miss}$, the transverse thrust value $\mathcal{T}$ calculated with \antikt jets of radius 0.4, the multiplicity of those jets, the azimuthal angle between the closest jet and missing transverse momentum direction $|\phi_{min}|$, the leading jet \pT and the scalar sum of the \pT of all jets $H_T$. Again, excellent agreement between the two models is observed.

\begin{figure}[ht]
    \centering
    \includegraphics[width=0.3\linewidth]{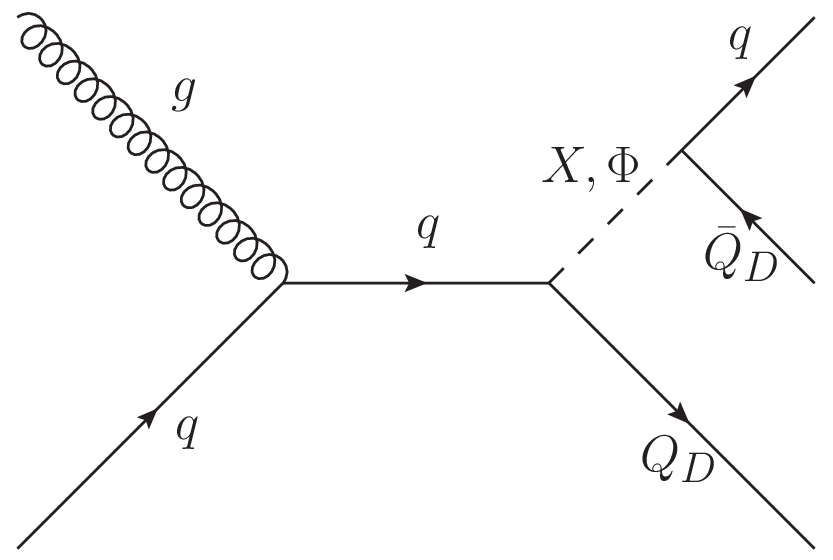}
    \hspace{2.5cm}
    \includegraphics[width=0.3\linewidth]{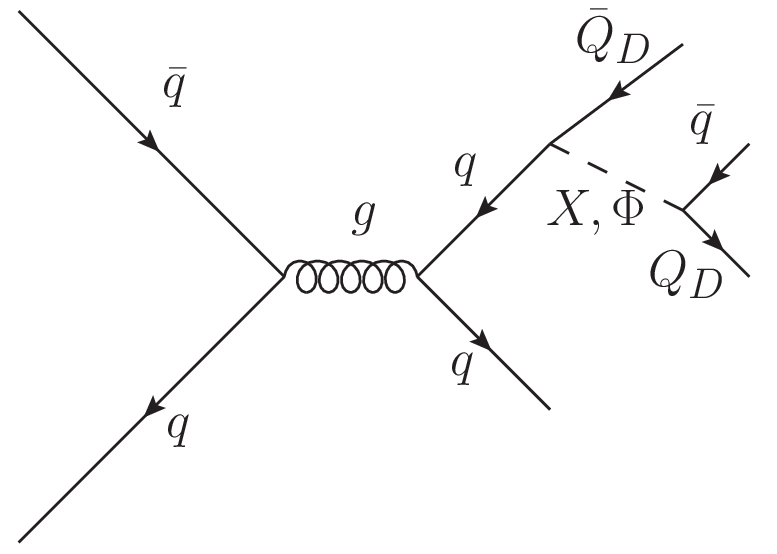}
    \caption{Example Feynman diagrams for the processes $q g \to Q_D \bar{Q}_D q$ and $q q \to Q_D \bar{Q}_D q q$ leading to the resonant enhancement for narrow mediator width.}
    \label{fig:res_enhancement}
\end{figure}

\begin{figure}[ht]
\centering
\includegraphics[width=0.45\textwidth]{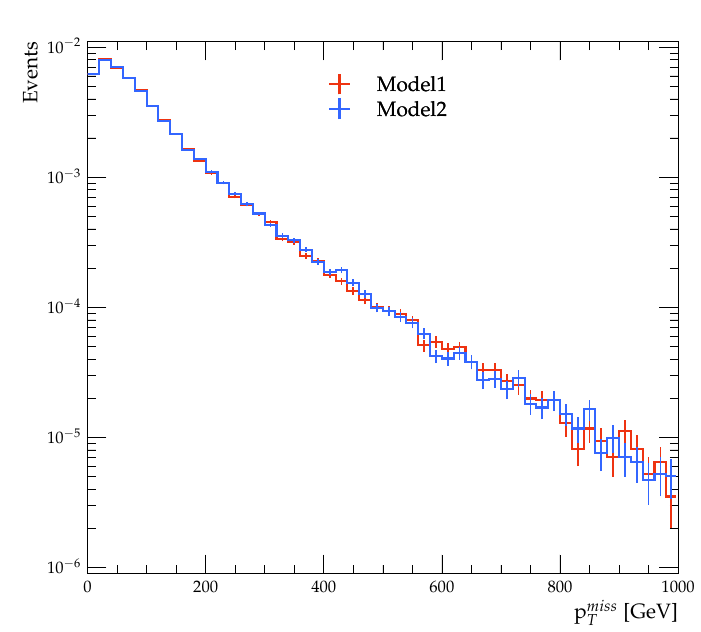}
\includegraphics[width=0.45\textwidth]{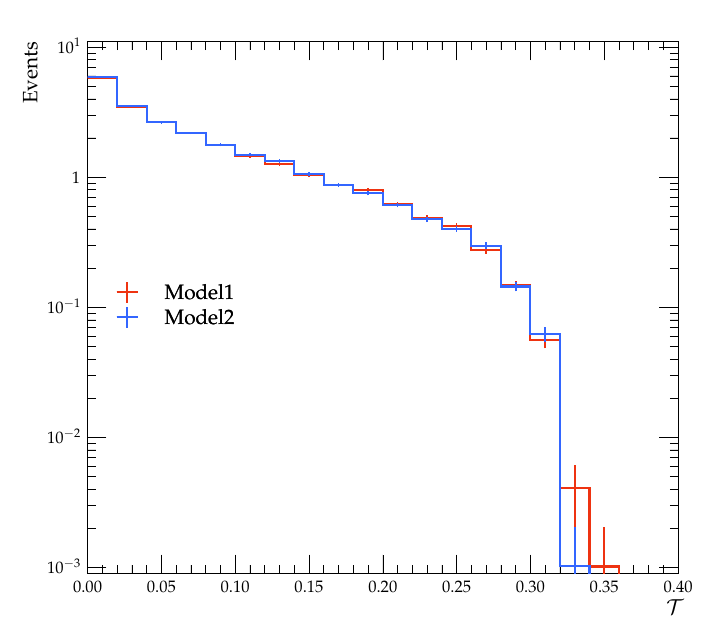}\\
\includegraphics[width=0.45\textwidth]{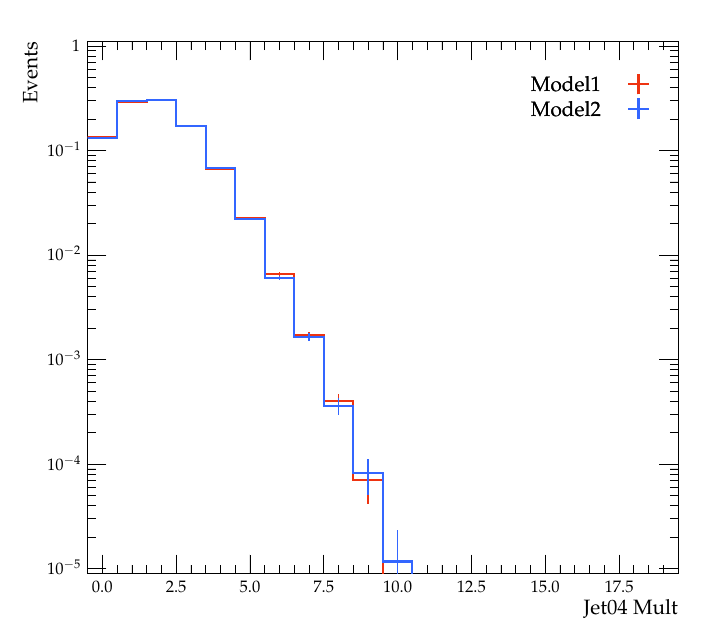 } 
\includegraphics[width=0.45\textwidth]{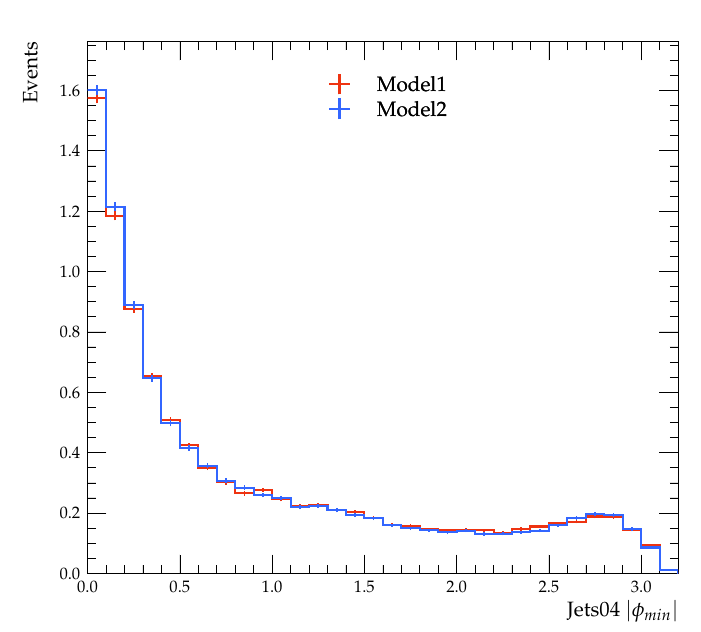}  \\
\includegraphics[width=0.45\textwidth]{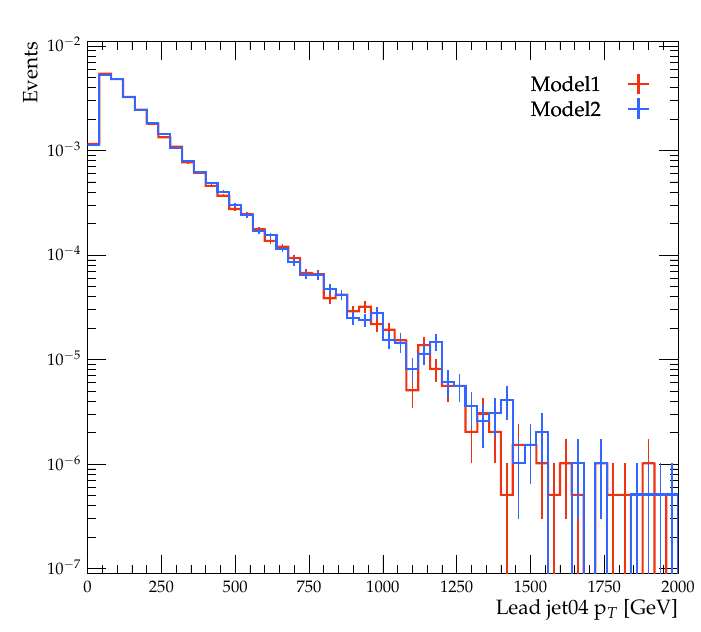}  
\includegraphics[width=0.45\textwidth]{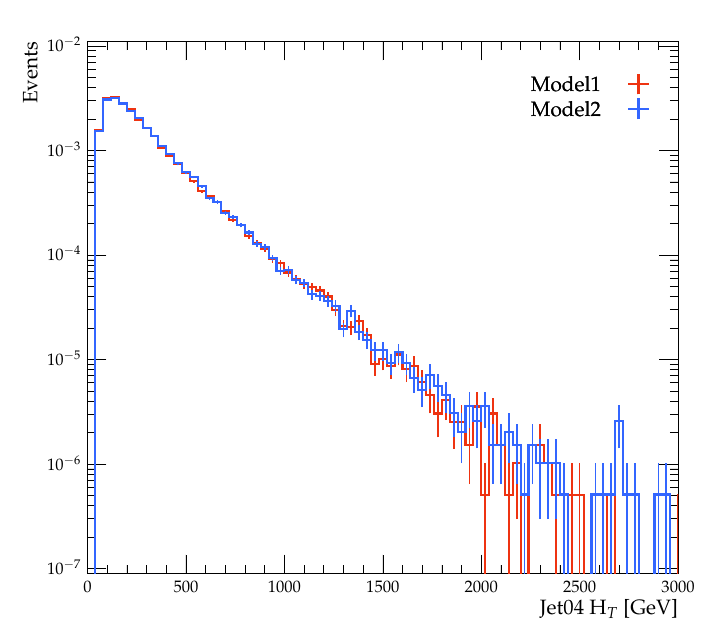}  
  \caption{Particle level Model1 against Model2 comparisons for the $p p \to Q_D \bar{Q}_D$ process with a mediator mass of $M=3000$~GeV and \Rinv\ fraction of 0.5, for a set of kinematic observables. Jet04 simply refers to the use of \antikt jets with radius parameter 0.4 to construct our observables.}
  \label{fig:comp-particle2to2}
\end{figure}

\begin{figure}[ht]
\centering
\includegraphics[width=0.45\textwidth]{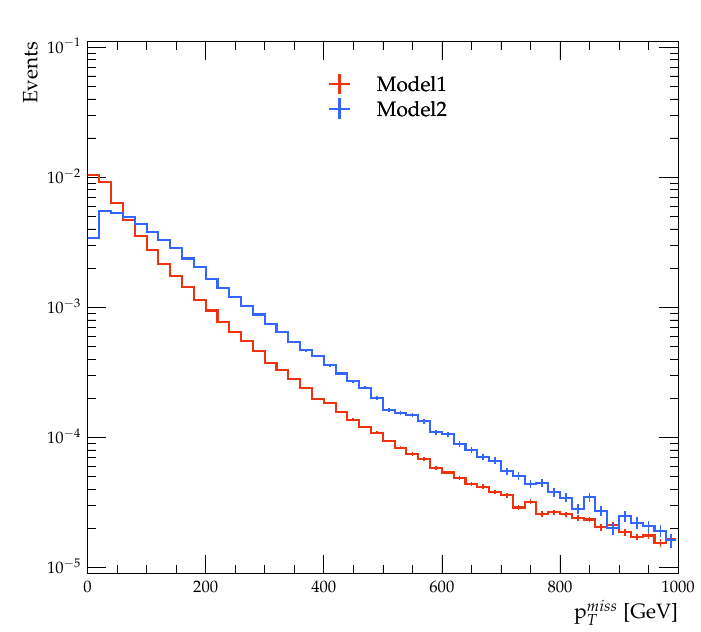}
\includegraphics[width=0.45\textwidth]{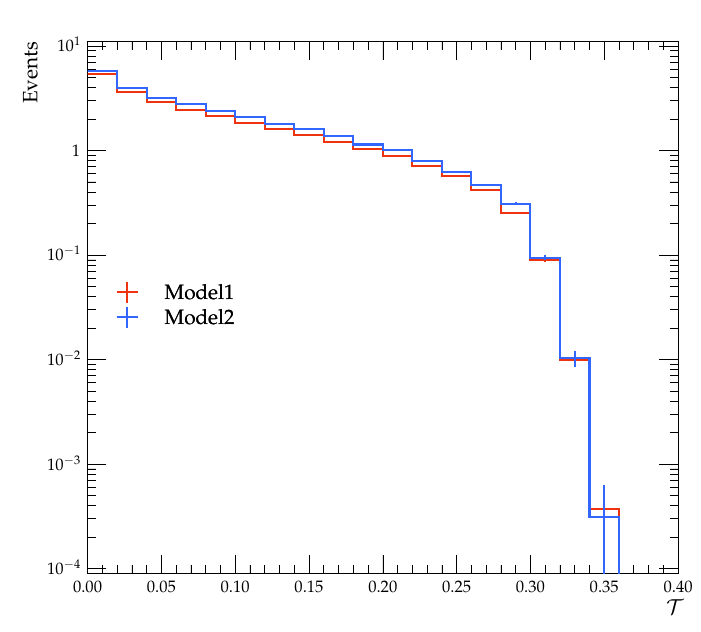}\\
\includegraphics[width=0.45\textwidth]{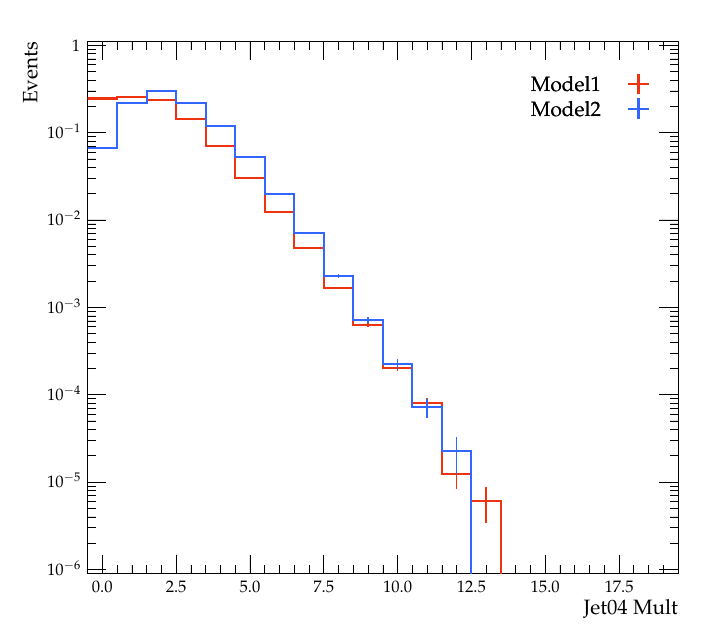 } 
\includegraphics[width=0.45\textwidth]{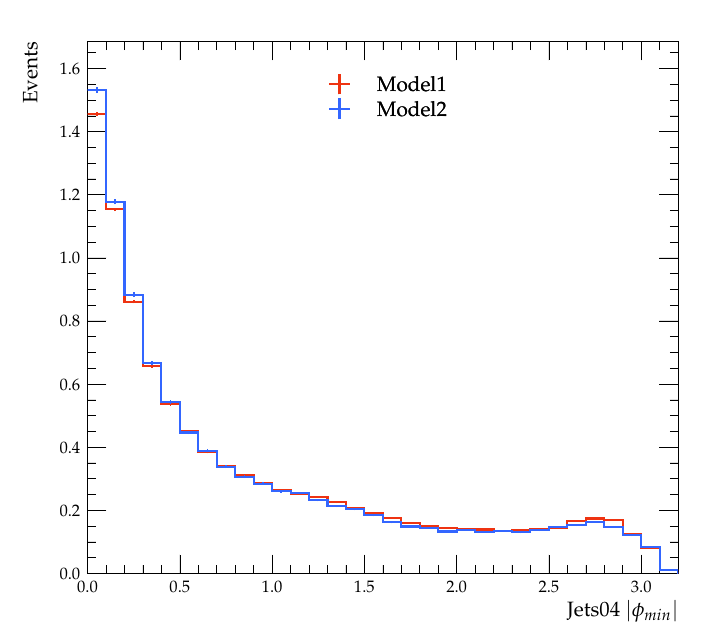}  \\
\includegraphics[width=0.45\textwidth]{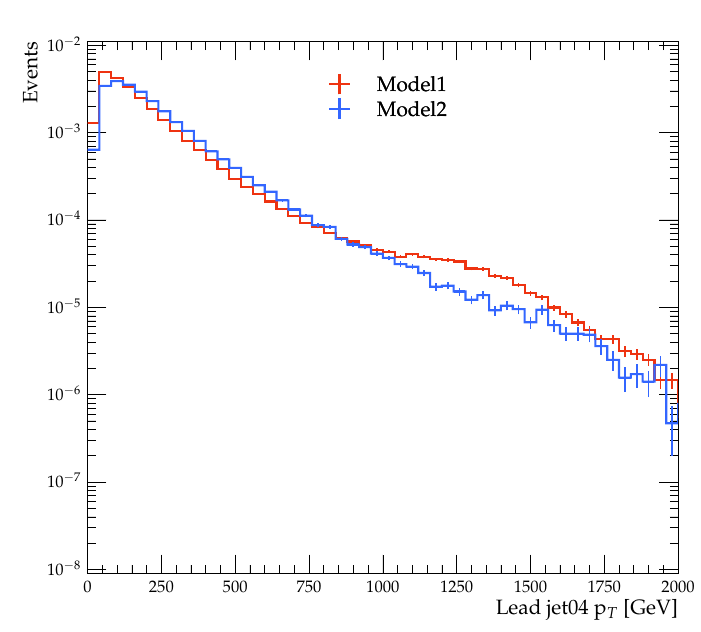}  
\includegraphics[width=0.45\textwidth]{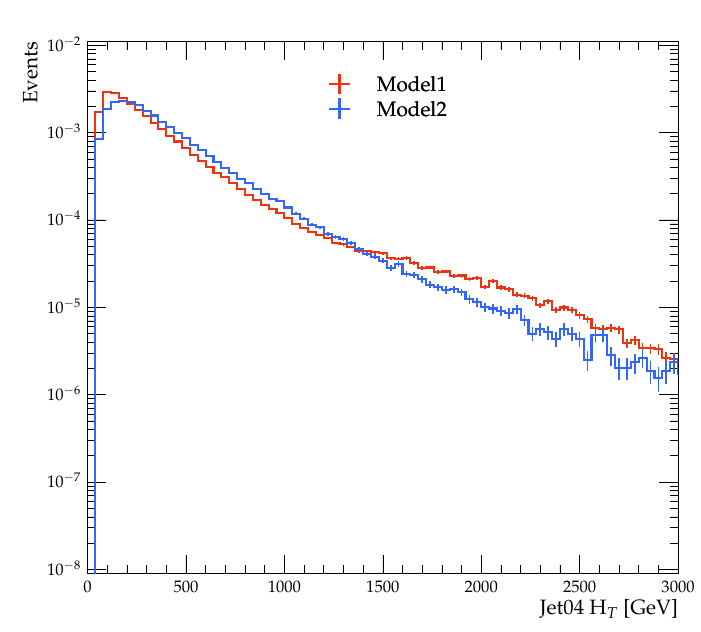}  
  \caption{Particle level Model1 against Model2 comparisons with up to two extra partons with a mediator mass of $M=3000$~GeV, with a \Rinv\ fraction of 0.5, for a set of kinematic observables. Jet04 simply refers to the use of \antikt jets with radius parameter 0.4 to construct our observables.}
  \label{fig:comp-particleFull}
\end{figure}

\paragraph{$\textbf{Full process}$}
Then, moving to the full process with up to two extra jets, we show the same kinematic distributions in Fig.~\ref{fig:comp-particleFull}.  
Note that the minimum \pT requirement for dark quarks was set at 200~GeV here, while all other parameters and cuts remain the same as before. 
Different from the parton and particle level comparison of the $2\to2$ process, some notable differences can be seen for the full process. In particular, a bump is observed in the \pT and $H_T$ distributions of Model1, that is not (or only very mildly) visible for Model2. Similarly, the $\pT^{miss}$ distributions show a notable difference with Model1 featuring a sharper and larger peak at smaller values and falling of faster than the distribution of Model2. Finally, the multiplicity distribution for Model1 is peaked at lower multiplicities and is broader than the one for Model2. The transverse thrust and the azimuthal angle between the closest jet and missing transverse momentum direction, on the other hand, are in good agreement.

The different features can be explained, when carefully considering the different properties of the models and the sub-processes contributing to the distributions. In Model1 each mediator has only one decay mode $\Phi_{bi}\to q_{i} \bar{\chi}_b$, while in Model2 the single mediator has nine possible decay modes $X\to Q_\alpha \bar{q}_i$ with $\alpha=1,2,3$ and $i = 1,2,3$. Thus, the decay widths of the mediator(s) in the two models are given as
\begin{align}
    \Gamma_{\Phi_{bj}} & = \frac{1}{16\pi}\lambda^2 m_\Phi \,, \\
    \Gamma_X &= 3\times N_\alpha \frac{1}{16\pi}\kappa^2 m_X\,,
\end{align}
where $N_\alpha$ is the number of dark quark flavors with non-zero couplings to SM quarks. For $M=3000$~GeV, $\kappa=\lambda = 1$ and $N_\alpha=3$ we get $\Gamma_\Phi\approx60$~GeV and $\Gamma_X\approx537$~GeV. The same values are obtained when calculating the decay widths in Madgraph. 
The narrow width in Model1 leads to a resonant enhancement: mediators produced in the s-channel dominantly have momentum around the mediator mass. The decay products of these mediators have a momentum of about half the mediator mass, leading to the bump. While this effect is in principle also present in Model2, the larger width smears out the resonance making it barely visible in the momentum distribution of the full process. Processes with a mediator in the s-channel only arise once one or more additional jets are allowed in the final state, more precisely in the processes $q g \to Q_D \bar{Q}_D q$ and $q q \to Q_D \bar{Q}_D q q$, where $q$ are SM quarks and $g$ gluons as shown in Fig.~\ref{fig:res_enhancement}.
We explicitly checked that the jets contributing to the bump originate from resonant mediator decays. 

This resonant enhancement does not only explain the bump in the \pT of the hardest jet and $H_T$ distributions, but also in the $\pT^{miss}$ distribution. In Model1 jets in the bump region carry a larger amount of energy away than in Model2, leading to $\pT^{miss}$ in Model1 being peaked at smaller energies and falling off faster. Similarly, the sharper peak at low \pT and the bump in the \pT distribution explains the different shapes of the multiplicity distribution: more low multiplicity events originate from the low \pT region and more high multiplicity events from the bump region.

\FloatBarrier

\section{Moving to a more efficient matching setup}
\label{sec:matching}

Previously the MLM~\cite{Mangano:2001xp, Mangano:2006rw} 
jet matching scheme with the matching parameter set to 100~GeV was used for all SVJ studies, following~\cite{Cohen:2017pzm}. However MLM matching
in general is not very efficient, as it discards the events which fail the matching criteria. 
So  a more modern and efficient CKKW-L~\cite{Catani:2001cc,Hoeche:2009rj,Platzer:2012bs,Lonnblad:2001iq, Lonnblad:2011xx} 
matching scheme was studied for Model2.~\footnote{By using: icckw$=0$ in Madgraph5, Merging:Process $= pp>{ddark1,4900101}{ddark1~,4900101}$ and Merging:nJetMax $= 2$ in Pythia\,8.} 
Again, the same representative value of mediator mass of 3000 GeV, with a \Rinv\ fraction of 0.5 was used.
Since the full generation enforced a minimum \pT requirement for dark quarks at 200~GeV here, the Madgraph merging scale needs to be at least 200 GeV for consistency. In Tab.~\ref{tab:matching} three different \textsc{Merging:TMS} values in Pythia\,8 \textsc{Main164} program are shown, comparing cross-sections and matching efficiency. While no dramatic difference was observed, the Pythia\,8
setting of \textsc{Merging:TMS} parameter at 250 GeV was preferred, as typically a slightly higher shower matching scale than that set in matrix element generator is preferable. The matching efficiency increases to 94\%, compared to an efficiency of about 70\% with MLM matching for this mediator mass.  

Two other important parameters in Pythia\,8 are \textsc{SpaceShower:rapidityOrder} and \textsc{SpaceShower:pTmaxMatch}. The former, when on, force the emissions after the first to be ordered in rapidity while the latter sets how the maximum shower evolution scale is set to match the scale of the hard process itself. No visible difference in the kinematic distributions was seen by turning the \textsc{SpaceShower:rapidityOrder} on or off, so it was decided to keep the default setting of on. For \textsc{SpaceShower:pTmaxMatch}, option 2, or the so called wimpy showers was chosen, which uses 
the factorization scale for an internal process and the scale value from Les Houches Events file. This by design strictly avoids double-counting, which is preferred.

\begin{figure}[ht]
\centering
\includegraphics[width=0.44\textwidth]{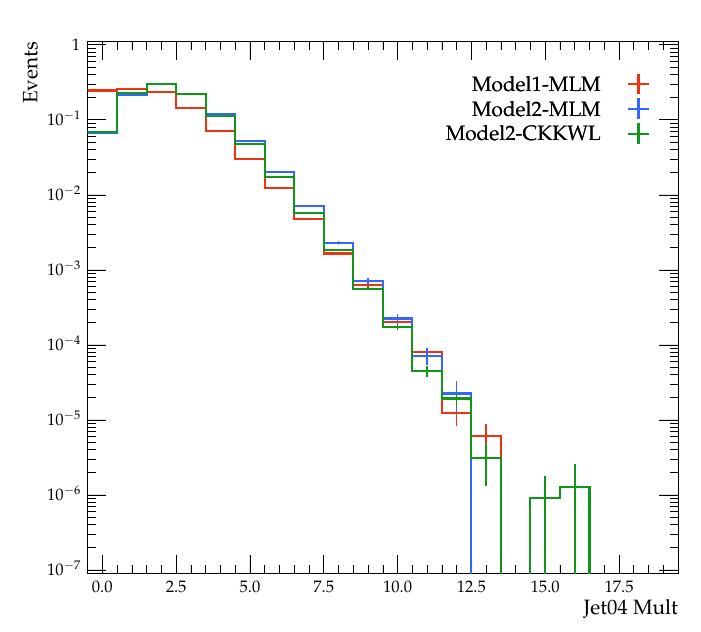}
\includegraphics[width=0.44\textwidth]{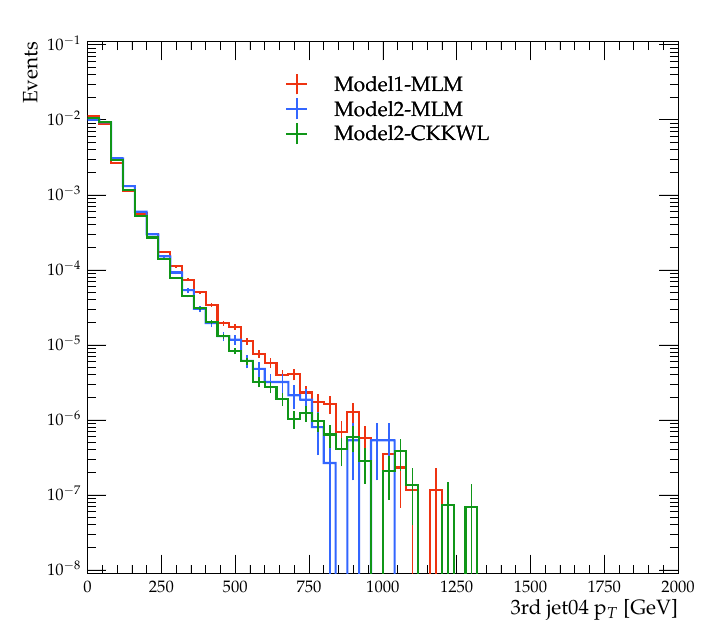} \\
\includegraphics[width=0.44\textwidth]{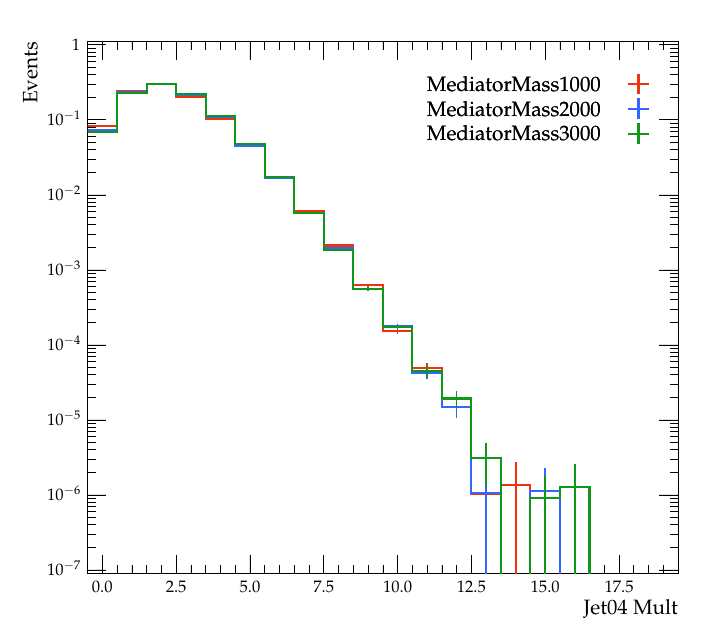} 
\includegraphics[width=0.44\textwidth]{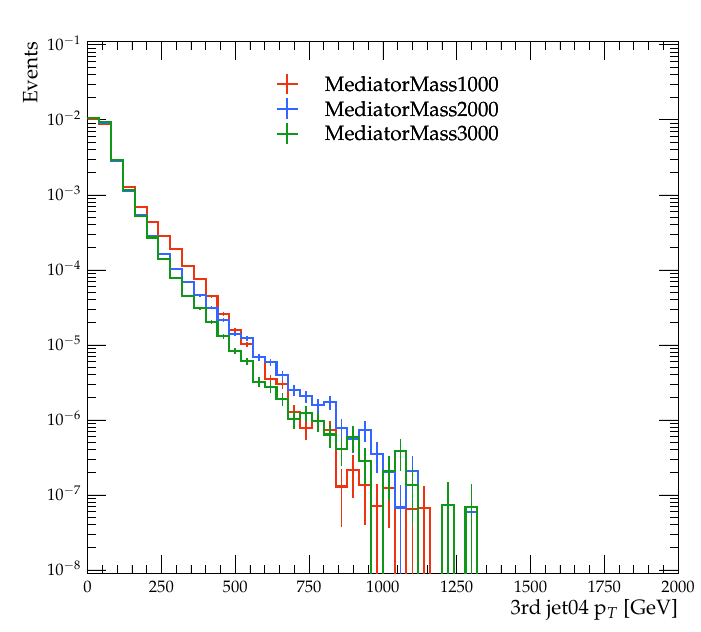} 
\caption{Comparison of matching setups for jet multiplicity and third jet \pT for an intermediate \Rinv\ value.}
\label{fig:comp-matching1}
\end{figure}

In Fig.~\ref{fig:comp-matching1} the distributions of jet multiplicity and the third jet \pT is shown, comparing first between Model1 with MLM and Model2 with both MLM and CKKW-L matching for the fixed mediator mass of 3000 GeV, and then just for Model2 with CKKW-L for three mediator masses, using \Rinv\ of 0.5 in both cases. The third jet \pT typically is the most sensitive to matching, as well as the jet multiplicity. As it can be seen, the difference is only visible when comparing the two different models, as was already seen in Fig.~\ref{fig:comp-particleFull}. The third jet \pT distribution is smooth, suggesting the matching is sane.

\begin{figure}[ht]
\centering
\includegraphics[width=0.44\textwidth]{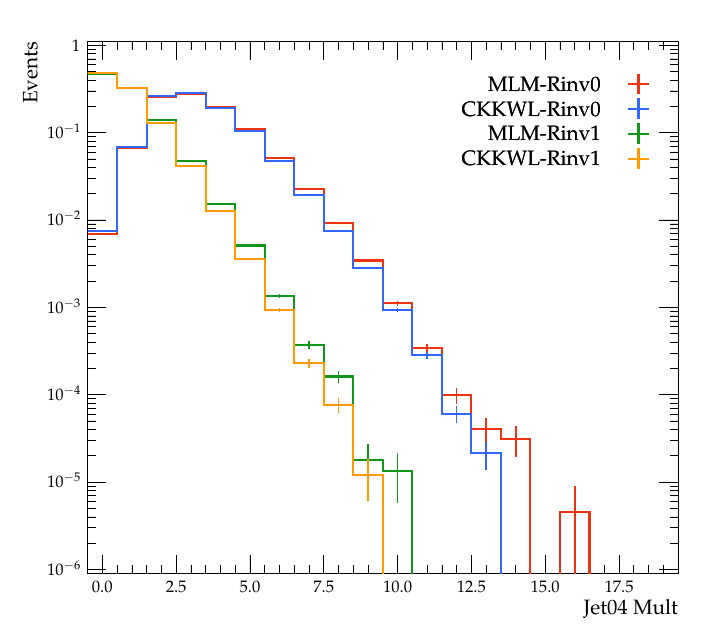}
\includegraphics[width=0.44\textwidth]{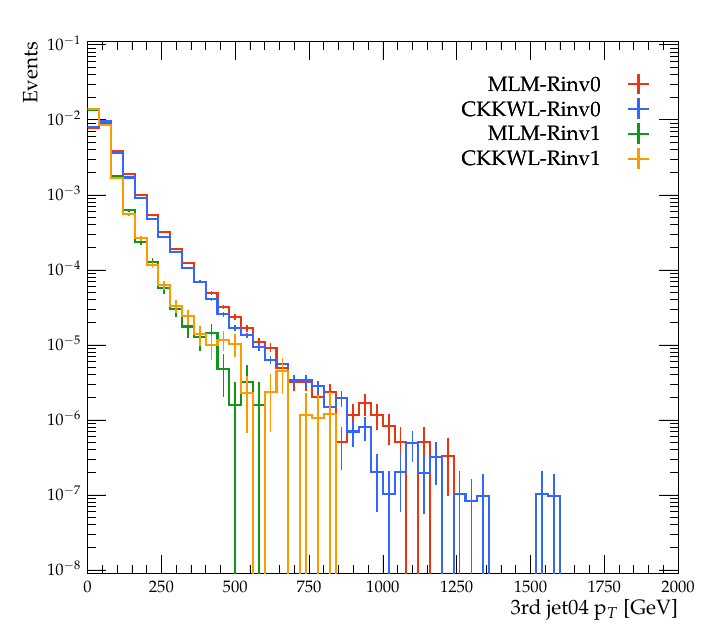} \\
\caption{Comparison of matching setups for jet multiplicity and third jet \pT
for extreme \Rinv\ values.}
\label{fig:comp-matching2}
\end{figure}

Another check is to use extreme values of \Rinv\,, zero and unity. The former indicates no dark hadrons in final state, so it is a pure SM-like multijet final state, while the latter will result in fully invisible decay of the dark quarks, so a less jetty pure SM final state as well. The same distributions as before are shown in Fig.~\ref{fig:comp-matching2},  and again except for some statistical fluctuations at very high values of jet \pT, the matching seems to be working.

\begin{table}[t]
\centering
{
\begin{tabular}{ccccc}
\toprule
MG5 ptlund [GeV] & MG5 $\sigma$ [pb] & Pythia\,8 TMS & Pythia $\sigma [pb]$ & Matching efficiency \\
\midrule
200 & 0.03769 & 200 & 0.03576 & 0.94 \\
200 & 0.03769 & 250 & 0.03567 & 0.94 \\
200 & 0.03769 & 300 & 0.03556 & 0.94 \\
\bottomrule
\end{tabular}
\caption{Effect of varying matching scale on cross-section and matching efficiency for a mediator mass of 3000 GeV and \Rinv\ of 0.5 for Model with CKKWL setup.}
\label{tab:matching}
}
\end{table}

\FloatBarrier

\section{Reinterpretation using the Run-2 ATLAS result}
\label{sec:reint}

\begin{figure}[ht]
\centering
\includegraphics[width=0.42\textwidth]{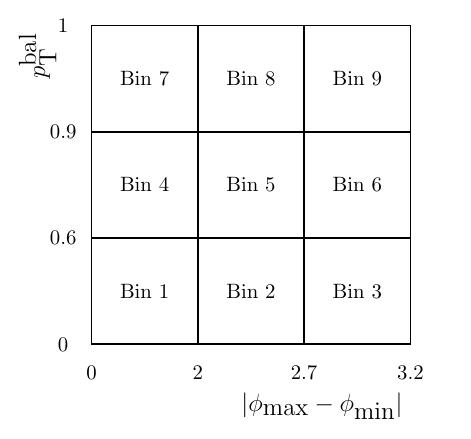}
\caption{Kinematic definition of the 9-bin grid used in ATLAS analysis of non-resonant production of semi-visible jets~\cite{ATLAS:2023swa}.}
\label{fig:9bin}
\end{figure}

The final study was to check if the existing experimental results constrain the proposed setup. A reinterpretation study was performed using the ATLAS result on t-channel semi-visible jet production ~\cite{ATLAS:2023swa}. The key distributions of the ATLAS analysis is the 9-bin fit shown in Fig. 3(a) in the cited reference. This 9-bin distribution is constructed from two uncorrelated variables, namely, the $p_T$ balance between the jets closest ($j_1$) and farthest ($j_2$) in azimuth from the $p^{miss}_T$ direction, and the azimuthal separation between $j_1$ and $j_2$. The definition of the 9-bin grid is shown in Fig.~\ref{fig:9bin}. The reinterpretation study was performed using the publicly available analysis code provided by the ATLAS experiment, in the Rivet software package repository~\cite{Bierlich:2024vqo}.
The analysis uses the full Run 2 ATLAS dataset of 139 fb$^{-1}$ collected at the center-of-mass energy of 13 TeV and the data is made available in HEPData~\cite{Maguire:2017ypu}. The analyses selections were implemented in the particle level, with available
smearing for jets applied~\cite{Buckley:2019stt}.

\begin{figure}[ht]
\centering
\includegraphics[width=0.44\textwidth]{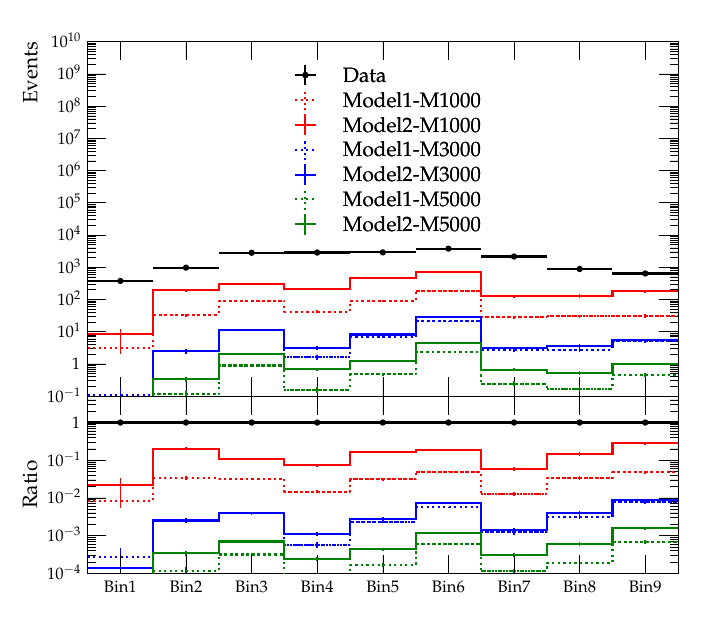}
\includegraphics[width=0.44\textwidth]{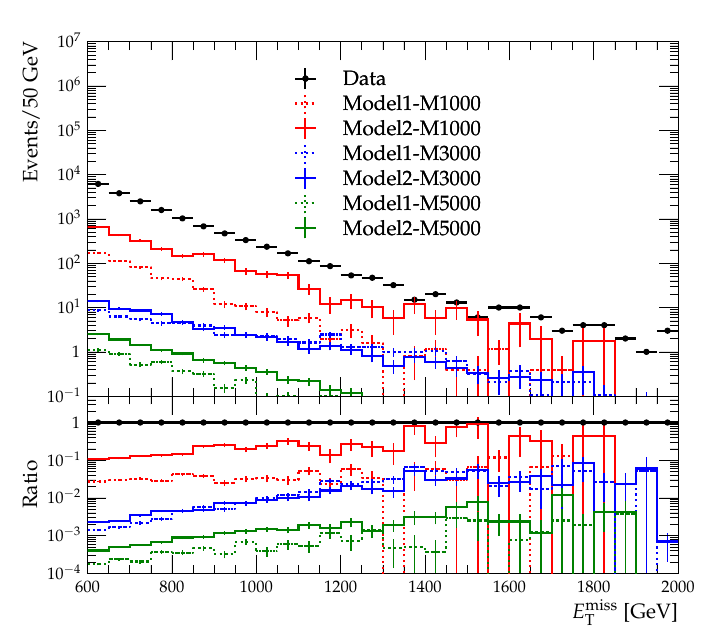}
\caption{Comparison of Model1 and Model2 results for three mediator masses with ATLAS data from Ref.~\cite{ATLAS:2023swa}.}
\label{fig:reint}
\end{figure}

The study is conducted with three mediator masses of 1000 GeV, 2000 GeV and 3000 GeV for \Rinv\ value of 0.3, assuming nominal coupling values for both the models.
Fig.~\ref{fig:reint} shows the 9-bin distribution for the three mediators masses along with the ratio of each to the data as well as the \metval distribution. None of the three mediator masses is seen to be excluded, and the ratios indicate by what factor the cross-section is below data, thereby indirectly suggesting what coupling values are excluded as well. From the \metval distribution, it is evident that the ATLAS search is more sensitive for higher values, keeping the lower \metval values very much in consideration.

A comment on the ATLAS result is warranted here. Although the ATLAS analysis employed a version of Model1 for signal generation, the generation settings differ from those adopted here. This is because subsequent studies~\cite{Albouy:2022cin} found the parameter choices used in the ATLAS analysis to be theoretically less well-motivated than those considered here. This explains why Model1 is not excluded in the present analysis, as can be seen from Fig.~\ref{fig:reint}. Nevertheless, since the ATLAS result employed a purely data-driven background estimation, its reinterpretation remains independent of the specific signal model choice. The CMS collaboration made their Run-2 non-resonant semi-visible jet analysis public recently~\cite{CMS-PAS-EXO-24-010}, however, the same study cannot be conducted using that result due to the machine-learning based analysis approach.

\FloatBarrier

\section{Summary}
\label{sec:summ}
Semi-visible jets are an exciting signature, which are motivated by a broad class of hidden valley and dark sector models, and are now part of the LHC new physics search program.  Concrete and well motivated benchmark models are required both for event generation and for consistent interpretation of the results of these searches. 

Here we compare two benchmark models for t-channel SVJ production: the original SVJ model of~\cite{Cohen:2015toa} (Model1) which features 24 individual particles as mediators, each of which only couples to a specific combination of SM and dark quarks, and a model with a single mediator based on the sneaky DM scenario~\cite{Carmona:2024tkg}, with a more complex coupling structure between quarks and dark quarks (Model2). The motivation for this comparison is rather pragmatic: The UFO implementation of Model1 is not compatible with the most recent versions of the Pythia\,8 event generator, which significantly updated the simulation and modeling of hidden valleys and dark showers. Since Model2 also features a t-channel mediator, the same types of Feynman diagrams appear in the event generation, suggesting that the corresponding UFO model could alternatively be used. 

Our main results are the following. First, Model2 is suitable for generating events for future SVJ searches, which we demonstrate by a careful comparison of kinematic distributions both at the parton level and at the level of hadronized events. Due to the different number of mediators, the pair production cross sections differ from each other for the same value of the couplings, which is well approximated by the ratio of the number of subprocesses that contribute. For the full production process there are contributions which scale with different powers of the coupling, leading to additional variations in the cross sections between the models, as shown in Tab.~\ref{tab:xsec}. The main differences in the distributions are a more pronounced bump in the leading jet $p_T$ in Model1 at half the mediator mass and a slightly harder $p_T^{\rm miss}$ distribution for Model2. Both differences can be traced back to a significantly larger mediator width in Model2 due to the larger number of final states. 

Overall the distributions are sufficiently close such that the current SVJ search strategies remain sensitive. We therefore perform a recast of the ATLAS SVJ search using the public Rivet implementation. This is actually done for both Model1 and Model2, since the theory recommendations for the parameters for Model1 were updated in a recent study.
The search is not sensitive yet for $\lambda=\kappa=1$ Yukawa couplings, but excludes $\lambda> 2.1 \; (\kappa>1.4)$ for Model1 (Model2) and mediator masses of 1~TeV. Both models show a very similar behaviour of the sensitivity in the different search bins, further supporting the statement that Model2 is a good benchmark for future searches. 

The results also show good prospects to probe the model up to several TeV mediator masses with future high-luminosity searches. Some improvements can also be expected by optimizing the search for the slightly modified kinematic distributions. Another promising avenue is to exploit long lived particle signatures from the unstable dark pions in the dark showers, leading to emerging semi-visible jets~\cite{Carmona:2024tkg,Carrasco:2025bct}. 

Finally, as discussed in Sec.~\ref{sec:matching}, a proposal to use more efficient CKKW-L matching is put forward. This significantly improves the matching efficiency, while leaving no statistically significant changes in the kinematic distributions.


\section*{Acknowledgments}

We dedicate this paper to the memory of our colleague and friend, Deepak Kar, who passed away during its completion. Deepak was instrumental in initiating this project and produced many of the results presented here.
DK thanks the generous support by Wolfson Foundation and Royal Society to allow him to spend his sabbatical year at the University of Glasgow in 2024 and Competitive Programme for Rated Researchers grant from National Research Foundation, South Africa (Grant Number:~CPRR240315209380). PS acknowledges support by the Cluster of Excellence “Precision Physics, Fundamental Interactions, and
Structure of Matter” (PRISMA++ EXC 2118/2) funded
by the Deutsche Forschungsgemeinschaft (DFG, German
Research Foundation), (Project No. 390831469). The work of CS is supported by the Office of High Energy Physics of the U.S. Department of Energy
under contract DE-AC02-05CH11231.
SS is supported by European Research Council grant REALDARK (Grant Agreement no:~101002463). 
We would like to thank Tim Cohen, Matt Strassler and Wrishik Naskar for
useful inputs. The discussions in the recent dark showers workshop at CERN contributed to finalising the work.

\bibliographystyle{elsarticle-num.bst}
\bibliography{ref.bib}

@article{Catani:2001cc,
    author = "Catani, S. and Krauss, F. and Kuhn, R. and Webber, B. R.",
    title = "{QCD matrix elements + parton showers}",
    eprint = "hep-ph/0109231",
    archivePrefix = "arXiv",
    reportNumber = "CERN-TH-2000-367, CAVENDISH-HEP-00-03",
    doi = "10.1088/1126-6708/2001/11/063",
    journal = "JHEP",
    volume = "11",
    pages = "063",
    year = "2001"
}

@article{Hoeche:2009rj,
    author = "Hoeche, Stefan and Krauss, Frank and Schumann, Steffen and Siegert, Frank",
    title = "{QCD matrix elements and truncated showers}",
    eprint = "0903.1219",
    archivePrefix = "arXiv",
    primaryClass = "hep-ph",
    reportNumber = "ZU-TH-02-09, IPPP-09-14, DCPT-09-28, HD-THEP-09-2, MCNET-09-04",
    doi = "10.1088/1126-6708/2009/05/053",
    journal = "JHEP",
    volume = "05",
    pages = "053",
    year = "2009"
}

@article{Platzer:2012bs,
    author = {Pl{\"a}tzer, Simon},
    title = "{Controlling inclusive cross sections in parton shower + matrix element merging}",
    eprint = "1211.5467",
    archivePrefix = "arXiv",
    primaryClass = "hep-ph",
    reportNumber = "DESY-12-215, MCNET-12-13",
    doi = "10.1007/JHEP08(2013)114",
    journal = "JHEP",
    volume = "08",
    pages = "114",
    year = "2013"
}

@article{Carrasco:2025bct,
    author = "Carrasco, Juliana and Kulkarni, Suchita and Liu, Wei and Lockyer, Joshua and Zurita, Jose",
    title = "{Semi-visible emerging jets}",
    eprint = "2511.02918",
    archivePrefix = "arXiv",
    primaryClass = "hep-ph",
    doi = "10.1007/JHEP07(2026)219",
    journal = "JHEP",
    volume = "07",
    pages = "219",
    year = "2026"
}

@article{Schwaller:2015gea,
    author = "Schwaller, Pedro and Stolarski, Daniel and Weiler, Andreas",
    title = "{Emerging Jets}",
    eprint = "1502.05409",
    archivePrefix = "arXiv",
    primaryClass = "hep-ph",
    reportNumber = "CERN-PH-TH-2015-031, DESY-15-026",
    doi = "10.1007/JHEP05(2015)059",
    journal = "JHEP",
    volume = "05",
    pages = "059",
    year = "2015"
}

@article{Bertone:2018krk,
    author = "Bertone, Gianfranco and Tait, Tim, M.P.",
    title = "{A new era in the search for dark matter}",
    eprint = "1810.01668",
    archivePrefix = "arXiv",
    primaryClass = "astro-ph.CO",
    doi = "10.1038/s41586-018-0542-z",
    journal = "Nature",
    volume = "562",
    number = "7725",
    pages = "51--56",
    year = "2018"
}

@article{Bernreuther:2019pfb,
    author = "Bernreuther, Elias and Kahlhoefer, Felix and Krämer, Michael and Tunney, Patrick",
    title = "{Strongly interacting dark sectors in the early Universe and at the LHC through a simplified portal}",
    eprint = "1907.04346",
    archivePrefix = "arXiv",
    primaryClass = "hep-ph",
    reportNumber = "TTK-19-25, P3H-19-019",
    doi = "10.1007/JHEP01(2020)162",
    journal = "JHEP",
    volume = "01",
    pages = "162",
    year = "2020"
}

@article{Beauchesne:2017yhh,
    author = "Beauchesne, Hugues and Bertuzzo, Enrico and Grilli Di Cortona, Giovanni and Tabrizi, Zahra",
    title = "{Collider phenomenology of Hidden Valley mediators of spin 0 or 1/2 with semivisible jets}",
    eprint = "1712.07160",
    archivePrefix = "arXiv",
    primaryClass = "hep-ph",
    doi = "10.1007/JHEP08(2018)030",
    journal = "JHEP",
    volume = "08",
    pages = "030",
    year = "2018"
}

@article{Park:2017rfb,
    author = "Park, Myeonghun and Zhang, Mengchao",
    title = "{Tagging a jet from a dark sector with Jet-substructures at colliders}",
    eprint = "1712.09279",
    archivePrefix = "arXiv",
    primaryClass = "hep-ph",
    doi = "10.1103/PhysRevD.100.115009",
    journal = "Phys. Rev. D",
    volume = "100",
    number = "11",
    pages = "115009",
    year = "2019"
}

@article{Mies:2020mzw,
    author = "Mies, Hanna and Scherb, Christiane and Schwaller, Pedro",
    title = "{Collider constraints on dark mediators}",
    eprint = "2011.13990",
    archivePrefix = "arXiv",
    primaryClass = "hep-ph",
    reportNumber = "MITP/20-068,P3H-20-069,TTK-20-41",
    doi = "10.1007/JHEP04(2021)049",
    journal = "JHEP",
    volume = "04",
    pages = "049",
    year = "2021"
}

@article{Cohen:2015toa,
    author = "Cohen, Timothy and Lisanti, Mariangela and Lou, Hou Keong",
    title = "{Semivisible Jets: Dark Matter Undercover at the LHC}",
    eprint = "1503.00009",
    archivePrefix = "arXiv",
    primaryClass = "hep-ph",
    doi = "10.1103/PhysRevLett.115.171804",
    journal = "Phys. Rev. Lett.",
    volume = "115",
    number = "17",
    pages = "171804",
    year = "2015"
}

@article{Cohen:2017pzm,
    author = "Cohen, Timothy and Lisanti, Mariangela and Lou, Hou Keong and Mishra-Sharma, Siddharth",
    title = "{LHC Searches for Dark Sector Showers}",
    eprint = "1707.05326",
    archivePrefix = "arXiv",
    primaryClass = "hep-ph",
    reportNumber = "PUPT-2529",
    doi = "10.1007/JHEP11(2017)196",
    journal = "JHEP",
    volume = "11",
    pages = "196",
    year = "2017"
}

@article{Cohen:2020afv,
    author = "Cohen, Timothy and Doss, Joel and Freytsis, Marat",
    archivePrefix = "arXiv",
    eprint = "2004.00631",
    month = "4",
    primaryClass = "hep-ph",
    title = "{Jet Substructure from Dark Sector Showers}",
    journal = "arXiv:2004.00631",
    year = "2020"
}

@article{Bernreuther:2020vhm,
    author = {Bernreuther, Elias and Finke, Thorben and Kahlhoefer, Felix and Kr\"amer, Michael and M\"uck, Alexander},
    title = "{Casting a graph net to catch dark showers}",
    eprint = "2006.08639",
    archivePrefix = "arXiv",
    primaryClass = "hep-ph",
    reportNumber = "TTK-20-17, P3H-20-025",
    doi = "10.21468/SciPostPhys.10.2.046",
    journal = "SciPost Phys.",
    volume = "10",
    number = "2",
    pages = "046",
    year = "2021"
}

@article{Maguire:2017ypu,
    author = "Maguire, Eamonn and Heinrich, Lukas and Watt, Graeme",
    editor = "Mount, Richard and Tull, Craig",
    title = "{HEPData: a repository for high energy physics data}",
    eprint = "1704.05473",
    archivePrefix = "arXiv",
    primaryClass = "hep-ex",
    reportNumber = "IPPP-17-31",
    doi = "10.1088/1742-6596/898/10/102006",
    journal = "J. Phys. Conf. Ser.",
    volume = "898",
    number = "10",
    pages = "102006",
    year = "2017"
}

@article{Buckley:2019stt,
    author = "Buckley, Andy and Kar, Deepak and Nordström, Karl",
    title = "{Fast simulation of detector effects in Rivet}",
    eprint = "1910.01637",
    archivePrefix = "arXiv",
    primaryClass = "hep-ph",
    reportNumber = "MCNET-19-22",
    doi = "10.21468/SciPostPhys.8.2.025",
    journal = "SciPost Phys.",
    volume = "8",
    pages = "025",
    year = "2020"
}

@article{Bierlich:2024vqo,
    author = "Bierlich, Christian and Buckley, Andy and Butterworth, Jonathan Mark and Gutschow, Christian and Lonnblad, Leif and Procter, Tomasz and Richardson, Peter and Yeh, Yoran",
    title = "{Robust independent validation of experiment and theory: Rivet version 4 release note}",
    eprint = "2404.15984",
    archivePrefix = "arXiv",
    primaryClass = "hep-ph",
    reportNumber = "MCNET-24-05",
    doi = "10.21468/SciPostPhysCodeb.36",
    journal = "SciPost Phys. Codeb.",
    volume = "36",
    pages = "1",
    year = "2024"
}

@article{Sjostrand:2014zea,
      author         = "Sjöstrand, Torbjörn and Ask, Stefan and Christiansen,
                        Jesper R. and Corke, Richard and Desai, Nishita and Ilten,
                        Philip and Mrenna, Stephen and Prestel, Stefan and
                        Rasmussen, Christine O. and Skands, Peter Z.",
      title          = "{An Introduction to PYTHIA 8.2}",
      journal        = "Comput. Phys. Commun.",
      volume         = "191",
      year           = "2015",
      pages          = "159-177",
      doi            = "10.1016/j.cpc.2015.01.024",
      eprint         = "1410.3012",
      archivePrefix  = "arXiv",
      primaryClass   = "hep-ph",
      reportNumber   = "LU-TP-14-36, MCNET-14-22, CERN-PH-TH-2014-190,
                        FERMILAB-PUB-14-316-CD, DESY-14-178, SLAC-PUB-16122",
      SLACcitation   = "%%CITATION = ARXIV:1410.3012;%%"
}

@article{Carloni:2011kk,
    author = "Carloni, Lisa and Rathsman, Johan and Sjostrand, Torbjorn",
    title = "{Discerning Secluded Sector gauge structures}",
    eprint = "1102.3795",
    archivePrefix = "arXiv",
    primaryClass = "hep-ph",
    reportNumber = "LU-TP-11-09, MCNET-11-06",
    doi = "10.1007/JHEP04(2011)091",
    journal = "JHEP",
    volume = "04",
    pages = "091",
    year = "2011"
}

@article{Mangano:2001xp,
      author         = "Mangano, Michelangelo L. and Moretti, Mauro and Pittau,
                        Roberto",
      title          = "{Multijet matrix elements and shower evolution in
                        hadronic collisions: $W b \bar{b}$ + $n$ jets as a case
                        study}",
      booktitle      = "{Workshop on Monte Carlo Generator Physics for Run II at
                        the Tevatron Batavia, Illinois, April 18-20, 2001}",
      journal        = "Nucl. Phys.",
      volume         = "B632",
      year           = "2002",
      pages          = "343-362",
      doi            = "10.1016/S0550-3213(02)00249-3",
      eprint         = "hep-ph/0108069",
      archivePrefix  = "arXiv",
      primaryClass   = "hep-ph",
      reportNumber   = "CERN-TH-2001-215",
      SLACcitation   = "%%CITATION = HEP-PH/0108069;%%"
}

@article{Lonnblad:2001iq,
    author = "Lonnblad, Leif",
    title = "{Correcting the color dipole cascade model with fixed order matrix elements}",
    eprint = "hep-ph/0112284",
    archivePrefix = "arXiv",
    reportNumber = "LU-TP-01-38",
    doi = "10.1088/1126-6708/2002/05/046",
    journal = "JHEP",
    volume = "05",
    pages = "046",
    year = "2002"
}

@article{Lonnblad:2011xx,
    author = "Lonnblad, Leif and Prestel, Stefan",
    title = "{Matching Tree-Level Matrix Elements with Interleaved Showers}",
    eprint = "1109.4829",
    archivePrefix = "arXiv",
    primaryClass = "hep-ph",
    doi = "10.1007/JHEP03(2012)019",
    journal = "JHEP",
    volume = "03",
    pages = "019",
    year = "2012"
}

@article{Cazzaniga:2022hxl,
    author = "Cazzaniga, Cesare and de Cosa, Annapaola",
    title = "{Leptons lurking in semi-visible jets at the LHC}",
    eprint = "2206.03909",
    archivePrefix = "arXiv",
    primaryClass = "hep-ph",
    doi = "10.1140/epjc/s10052-022-10775-2",
    journal = "Eur. Phys. J. C",
    volume = "82",
    number = "9",
    pages = "793",
    year = "2022"
}

@article{Albouy:2022cin,
    author = "Albouy, Guillaume and others",
    title = "{Theory, phenomenology, and experimental avenues for dark showers: a Snowmass 2021 report}",
    eprint = "2203.09503",
    archivePrefix = "arXiv",
    primaryClass = "hep-ph",
    reportNumber = "FERMILAB-PUB-22-303-PPD-SCD-T",
    doi = "10.1140/epjc/s10052-022-11048-8",
    journal = "Eur. Phys. J. C",
    volume = "82",
    number = "12",
    pages = "1132",
    year = "2022"
}

@article{Mangano:2006rw,
    author = "Mangano, Michelangelo L. and Moretti, Mauro and Piccinini, Fulvio and Treccani, Michele",
    title = "{Matching matrix elements and shower evolution for top-quark production in hadronic collisions}",
    eprint = "hep-ph/0611129",
    archivePrefix = "arXiv",
    reportNumber = "CERN-PH-TH-2006-232, FNT-T-2006-09",
    doi = "10.1088/1126-6708/2007/01/013",
    journal = "JHEP",
    volume = "01",
    pages = "013",
    year = "2007"
}

@article{Beauchesne:2018myj,
    author = "Beauchesne, Hugues and Bertuzzo, Enrico and Grilli Di Cortona, Giovanni",
    title = "{Dark matter in Hidden Valley models with stable and unstable light dark mesons}",
    eprint = "1809.10152",
    archivePrefix = "arXiv",
    primaryClass = "hep-ph",
    doi = "10.1007/JHEP04(2019)118",
    journal = "JHEP",
    volume = "04",
    pages = "118",
    year = "2019"
}

@article{Strassler:2006im,
    author = "Strassler, Matthew J. and Zurek, Kathryn M.",
    title = "{Echoes of a hidden valley at hadron colliders}",
    eprint = "hep-ph/0604261",
    archivePrefix = "arXiv",
    doi = "10.1016/j.physletb.2007.06.055",
    journal = "Phys. Lett. B",
    volume = "651",
    pages = "374--379",
    year = "2007"
}

@article{Renner:2018fhh,
    author = "Renner, Sophie and Schwaller, Pedro",
    title = "{A flavoured dark sector}",
    eprint = "1803.08080",
    archivePrefix = "arXiv",
    primaryClass = "hep-ph",
    doi = "10.1007/JHEP08(2018)052",
    journal = "JHEP",
    volume = "08",
    pages = "052",
    year = "2018"
}

@article{Manzari:2022iyn,
    author = "Manzari, Claudio Andrea and Profumo, Stefano",
    title = "{Flavor inspired model for dark matter}",
    eprint = "2206.06768",
    archivePrefix = "arXiv",
    primaryClass = "hep-ph",
    reportNumber = "ZU-TH 23/22, PSI-PR-22-15",
    doi = "10.1103/PhysRevD.106.075025",
    journal = "Phys. Rev. D",
    volume = "106",
    number = "7",
    pages = "075025",
    year = "2022"
}

@article{Strassler:2008fv,
    author = "Strassler, Matthew J.",
    title = "{On the Phenomenology of Hidden Valleys with Heavy Flavor}",
    eprint = "0806.2385",
    archivePrefix = "arXiv",
    primaryClass = "hep-ph",
    month = "6",
    year = "2008"
}

@article{Beauchesne:2022phk,
    author = "Beauchesne, Hugues and Cazzaniga, Cesare and de Cosa, Annapaola and Doglioni, Caterina and Fitschen, Tobias and di Cortona, Giovanni Grilli and Zhou, Ziyuan",
    title = "{Uncovering tau leptons-enriched semi-visible jets at the LHC}",
    eprint = "2212.11523",
    archivePrefix = "arXiv",
    primaryClass = "hep-ph",
    doi = "10.1140/epjc/s10052-023-11775-6",
    journal = "Eur. Phys. J. C",
    volume = "83",
    number = "7",
    pages = "599",
    year = "2023"
}

@article{ATLAS:2023swa,
    author = "{ATLAS Collaboration}",
    title = "{Search for non-resonant production of semi-visible jets using Run\textasciitilde{}2 data in ATLAS}",
    eprint = "2305.18037",
    archivePrefix = "arXiv",
    primaryClass = "hep-ex",
    reportNumber = "CERN-EP-2023-084",
    doi = "10.1016/j.physletb.2023.138324",
    journal = "Phys. Lett. B",
    volume = "848",
    pages = "138324",
    year = "2024"
}

@article{Han:2007ae,
    author = "Han, Tao and Si, Zongguo and Zurek, Kathryn M. and Strassler, Matthew J.",
    title = "{Phenomenology of hidden valleys at hadron colliders}",
    eprint = "0712.2041",
    archivePrefix = "arXiv",
    primaryClass = "hep-ph",
    reportNumber = "MADPH-07-1502, SDU-HEP-071201, RUNHETC-2007-31",
    doi = "10.1088/1126-6708/2008/07/008",
    journal = "JHEP",
    volume = "07",
    pages = "008",
    year = "2008"
}

@article{Bhardwaj:2024djv,
    author = "Bhardwaj, Akanksha and Englert, Christoph and Naskar, Wrishik and Ngairangbam, Vishal S. and Spannowsky, Michael",
    title = "{Equivariant, Safe and Sensitive -- Graph Networks for New Physics}",
    eprint = "2402.12449",
    archivePrefix = "arXiv",
    primaryClass = "hep-ph",
    reportNumber = "IPPP/24/07",
    doi = "10.1007/JHEP12(2024)105",
    journal = "JHEP",
    volume = "07",
    pages = "245",
    year = "2024"
}

@article{Batz:2023zef,
    author = "Batz, Austin and Cohen, Timothy and Curtin, David and Gemmell, Caleb and Kribs, Graham D.",
    title = "{Dark sector glueballs at the LHC}",
    eprint = "2310.13731",
    archivePrefix = "arXiv",
    primaryClass = "hep-ph",
    reportNumber = "CERN-TH-2023-194",
    doi = "10.1007/JHEP04(2024)070",
    journal = "JHEP",
    volume = "04",
    pages = "070",
    year = "2024"
}

@Article{10.21468/SciPostPhysCore.7.4.071,
	title={{2B or not 2B, a study of bottom-quark-philic semi-visible jets}},
	author={Deepak Kar and Wandile Nzuza and Sukanya Sinha},
	journal={SciPost Phys. Core},
	volume={7},
	pages={071},
	year={2024},
	publisher={SciPost},
	doi={10.21468/SciPostPhysCore.7.4.071},
	url={https://scipost.org/10.21468/SciPostPhysCore.7.4.071},
}

@article{Cohen:2023mya,
    author = "Cohen, Timothy and Roloff, Jennifer and Scherb, Christiane",
    title = "{Dark sector showers in the Lund jet plane}",
    eprint = "2301.07732",
    archivePrefix = "arXiv",
    primaryClass = "hep-ph",
    reportNumber = "CERN-TH-2023-007",
    doi = "10.1103/PhysRevD.108.L031501",
    journal = "Phys. Rev. D",
    volume = "108",
    number = "3",
    pages = "L031501",
    year = "2023"
}

@misc{dmsimp,
  title = {DMSimp model},
  howpublished = {\url{https://github.com/smsharma/SemivisibleJets}},
  note = {commit 4e606c4}
}

@misc{dmsneaky,
  title = {Sneaky DM model},
  howpublished = {\url{https://github.com/chscherb/t-channel_dark_QCD/}},
  note = {commit 8db614f}
}

@Article{Alwall:2014hca,
      author         = "Alwall, J. and Frederix, R. and Frixione, S. and Hirschi,
                        V. and Maltoni, F. and Mattelaer, O. and Shao, H. -S. and
                        Stelzer, T. and Torrielli, P. and Zaro, M.",
      title          = "{The automated computation of tree-level and
                        next-to-leading order differential cross sections, and
                        their matching to parton shower simulations}",
      journal        = "JHEP",
      volume         = "07",
      year           = "2014",
      pages          = "079",
      doi            = "10.1007/JHEP07(2014)079",
      eprint         = "1405.0301",
      archivePrefix  = "arXiv",
      primaryClass   = "hep-ph",
      reportNumber   = "CERN-PH-TH-2014-064, CP3-14-18, LPN14-066, MCNET-14-09,
                        ZU-TH-14-14",
      SLACcitation   = "%%CITATION = ARXIV:1405.0301;%%"
}

@article{Frederix:2012ps,
      author         = "Frederix, Rikkert and Frixione, Stefano",
      title          = "{Merging meets matching in MC@NLO}",
      journal        = "JHEP",
      volume         = "12",
      year           = "2012",
      pages          = "061",
      doi            = "10.1007/JHEP12(2012)061",
      eprint         = "1209.6215",
      archivePrefix  = "arXiv",
      primaryClass   = "hep-ph",
      reportNumber   = "CERN-PH-TH-2012-247, ZU-TH-21-12",
      SLACcitation   = "%%CITATION = ARXIV:1209.6215;%%"
}

@article{Carmona:2024tkg,
    author = "Carmona, Adrian and Elahi, Fatemeh and Scherb, Christiane and Schwaller, Pedro",
    title = "{Dark showers from sneaky dark matter}",
    eprint = "2411.15073",
    archivePrefix = "arXiv",
    primaryClass = "hep-ph",
    reportNumber = "MITP/24-083",
    doi = "10.1007/JHEP06(2025)198",
    journal = "JHEP",
    volume = "06",
    pages = "198",
    year = "2025"
}

@article{Kopp:2016yji,
    author = "Kopp, Joachim and Liu, Jia and Slatyer, Tracy R. and Wang, Xiao-Ping and Xue, Wei",
    title = "{Impeded Dark Matter}",
    eprint = "1609.02147",
    archivePrefix = "arXiv",
    primaryClass = "hep-ph",
    reportNumber = "MIT-CTP-4832, MITP-16-092",
    doi = "10.1007/JHEP12(2016)033",
    journal = "JHEP",
    volume = "12",
    pages = "033",
    year = "2016"
}

@article{Essig:2009nc,
    author = "Essig, Rouven and Schuster, Philip and Toro, Natalia",
    title = "{Probing Dark Forces and Light Hidden Sectors at Low-Energy e+e- Colliders}",
    eprint = "0903.3941",
    archivePrefix = "arXiv",
    primaryClass = "hep-ph",
    reportNumber = "SLAC-PUB-13561, SU-ITP-09-12",
    doi = "10.1103/PhysRevD.80.015003",
    journal = "Phys. Rev. D",
    volume = "80",
    pages = "015003",
    year = "2009"
}

@article{Bhattacharya:2013kma,
    author = "Bhattacharya, Subhaditya and Meli{\'c}, Bla{\v{z}}enka and Wudka, Jos{\'e}",
    title = "{Pionic Dark Matter}",
    eprint = "1307.2647",
    archivePrefix = "arXiv",
    primaryClass = "hep-ph",
    doi = "10.1007/JHEP02(2014)115",
    journal = "JHEP",
    volume = "02",
    pages = "115",
    year = "2014"
}

@article{Cline:2013zca,
    author = "Cline, James M. and Liu, Zuowei and Moore, Guy D. and Xue, Wei",
    title = "{Composite strongly interacting dark matter}",
    eprint = "1312.3325",
    archivePrefix = "arXiv",
    primaryClass = "hep-ph",
    doi = "10.1103/PhysRevD.90.015023",
    journal = "Phys. Rev. D",
    volume = "90",
    number = "1",
    pages = "015023",
    year = "2014"
}

@article{Hochberg:2014kqa,
    author = "Hochberg, Yonit and Kuflik, Eric and Murayama, Hitoshi and Volansky, Tomer and Wacker, Jay G.",
    title = "{Model for Thermal Relic Dark Matter of Strongly Interacting Massive Particles}",
    eprint = "1411.3727",
    archivePrefix = "arXiv",
    primaryClass = "hep-ph",
    doi = "10.1103/PhysRevLett.115.021301",
    journal = "Phys. Rev. Lett.",
    volume = "115",
    number = "2",
    pages = "021301",
    year = "2015"
}

@article{Harigaya:2016rwr,
    author = "Harigaya, Keisuke and Nomura, Yasunori",
    title = "{Light Chiral Dark Sector}",
    eprint = "1603.03430",
    archivePrefix = "arXiv",
    primaryClass = "hep-ph",
    doi = "10.1103/PhysRevD.94.035013",
    journal = "Phys. Rev. D",
    volume = "94",
    number = "3",
    pages = "035013",
    year = "2016"
}

@article{Berlin:2018pwi,
    author = "Berlin, Asher and Gori, Stefania and Schuster, Philip and Toro, Natalia",
    title = "{Dark Sectors at the Fermilab SeaQuest Experiment}",
    eprint = "1804.00661",
    archivePrefix = "arXiv",
    primaryClass = "hep-ph",
    reportNumber = "SLAC-PUB-17238",
    doi = "10.1103/PhysRevD.98.035011",
    journal = "Phys. Rev. D",
    volume = "98",
    number = "3",
    pages = "035011",
    year = "2018"
}

@article{Beauchesne:2019ato,
    author = "Beauchesne, Hugues and Grilli di Cortona, Giovanni",
    title = "{Classification of dark pion multiplets as dark matter candidates and collider phenomenology}",
    eprint = "1910.10724",
    archivePrefix = "arXiv",
    primaryClass = "hep-ph",
    doi = "10.1007/JHEP02(2020)196",
    journal = "JHEP",
    volume = "02",
    pages = "196",
    year = "2020"
}

@article{Contino:2020god,
    author = "Contino, Roberto and Podo, Alessandro and Revello, Filippo",
    title = "{Composite Dark Matter from Strongly-Interacting Chiral Dynamics}",
    eprint = "2008.10607",
    archivePrefix = "arXiv",
    primaryClass = "hep-ph",
    doi = "10.1007/JHEP02(2021)091",
    journal = "JHEP",
    volume = "02",
    pages = "091",
    year = "2021"
}

@article{Chu:2024rrv,
    author = "Chu, Xiaoyong and Nikolic, Marco and Pradler, Josef",
    title = "{Even SIMP miracles are possible}",
    eprint = "2401.12283",
    archivePrefix = "arXiv",
    primaryClass = "hep-ph",
    doi = "10.1103/PhysRevLett.133.021003",
    journal = "Phys. Rev. Lett.",
    volume = "133",
    number = "2",
    pages = "2",
    year = "2024"
}

@article{Garcia-Cely:2024ivo,
    author = "Garc{\'\i}a-Cely, Camilo and Landini, Giacomo and Zapata, {\'O}scar",
    title = "{Dark matter in QCD-like theories with a theta vacuum: Cosmological and astrophysical implications}",
    eprint = "2405.10367",
    archivePrefix = "arXiv",
    primaryClass = "hep-ph",
    doi = "10.1103/PhysRevD.111.063044",
    journal = "Phys. Rev. D",
    volume = "111",
    number = "6",
    pages = "063044",
    year = "2025"
}

@article{Maleknejad:2022gyf,
    author = "Maleknejad, Azadeh and McDonough, Evan",
    title = "{Ultralight pion and superheavy baryon dark matter}",
    eprint = "2205.12983",
    archivePrefix = "arXiv",
    primaryClass = "hep-ph",
    reportNumber = "CERN-TH-2022-086",
    doi = "10.1103/PhysRevD.106.095011",
    journal = "Phys. Rev. D",
    volume = "106",
    number = "9",
    pages = "095011",
    year = "2022"
}

@article{Alexander:2023wgk,
    author = "Alexander, Stephon and Gilmer, Humberto and Manton, Tucker and McDonough, Evan",
    title = "{{\ensuremath{\pi}}-axion and {\ensuremath{\pi}}-axiverse of dark QCD}",
    eprint = "2304.11176",
    archivePrefix = "arXiv",
    primaryClass = "hep-ph",
    doi = "10.1103/PhysRevD.108.123014",
    journal = "Phys. Rev. D",
    volume = "108",
    number = "12",
    pages = "123014",
    year = "2023"
}

@article{Alexander:2024nvi,
    author = "Alexander, Stephon and Manton, Tucker and McDonough, Evan",
    title = "{Field theory axiverse}",
    eprint = "2404.11642",
    archivePrefix = "arXiv",
    primaryClass = "hep-ph",
    doi = "10.1103/PhysRevD.109.116019",
    journal = "Phys. Rev. D",
    volume = "109",
    number = "11",
    pages = "116019",
    year = "2024"
}

@article{Linthorne:2021oiz,
    author = "Linthorne, Dylan and Stolarski, Daniel",
    title = "{Triggering on emerging jets}",
    eprint = "2103.08620",
    archivePrefix = "arXiv",
    primaryClass = "hep-ph",
    doi = "10.1103/PhysRevD.104.035019",
    journal = "Phys. Rev. D",
    volume = "104",
    number = "3",
    pages = "035019",
    year = "2021"
}

@article{Carmona:2021seb,
    author = "Carmona, Adrian and Scherb, Christiane and Schwaller, Pedro",
    title = "{Charming ALPs}",
    eprint = "2101.07803",
    archivePrefix = "arXiv",
    primaryClass = "hep-ph",
    reportNumber = "MITP-21-003",
    doi = "10.1007/JHEP08(2021)121",
    journal = "JHEP",
    volume = "08",
    pages = "121",
    year = "2021"
}

@article{Carmona:2022jid,
    author = "Carmona, Adrian and Elahi, Fatemeh and Scherb, Christiane and Schwaller, Pedro",
    title = "{The ALPs from the top: searching for long lived axion-like particles from exotic top decays}",
    eprint = "2202.09371",
    archivePrefix = "arXiv",
    primaryClass = "hep-ph",
    reportNumber = "MITP/22-017",
    doi = "10.1007/JHEP07(2022)122",
    journal = "JHEP",
    volume = "07",
    pages = "122",
    year = "2022"
}

@article{Jubb:2017rhm,
    author = "Jubb, Thomas and Kirk, Matthew and Lenz, Alexander",
    title = "{Charming Dark Matter}",
    eprint = "1709.01930",
    archivePrefix = "arXiv",
    primaryClass = "hep-ph",
    reportNumber = "IPPP-17-64",
    doi = "10.1007/JHEP12(2017)010",
    journal = "JHEP",
    volume = "12",
    pages = "010",
    year = "2017"
}

@article{Agrawal:2014aoa,
    author = "Agrawal, Prateek and Blanke, Monika and Gemmler, Katrin",
    title = "{Flavored Dark Matter beyond Minimal Flavor Violation}",
    eprint = "1405.6709",
    archivePrefix = "arXiv",
    primaryClass = "hep-ph",
    reportNumber = "CERN-PH-TH-2014-098, FERMILAB-PUB-14-141-T",
    doi = "10.1007/JHEP10(2014)072",
    journal = "JHEP",
    volume = "10",
    pages = "072",
    year = "2014"
}

@article{Butterworth:2023cgz,
    author = "Butterworth, Jonathan and Cazzaniga, Cesare and Garcia-Bellido, Aran and Kar, Deepak and Kulkarni, Suchita and Schwaller, Pedro and Sinha, Sukanya and Wilson-Edwards, Danielle and Zurita, Jose",
    title = "{MITP Colours in Darkness workshop summary report}",
    eprint = "2311.16330",
    archivePrefix = "arXiv",
    primaryClass = "hep-ph",
    month = "11",
    year = "2023"
}

@article{Carrasco:2023loy,
    author = "Carrasco, Juliana and Zurita, Jos{\'e}",
    title = "{Emerging jet probes of strongly interacting dark sectors}",
    eprint = "2307.04847",
    archivePrefix = "arXiv",
    primaryClass = "hep-ph",
    doi = "10.1007/JHEP01(2024)034",
    journal = "JHEP",
    volume = "01",
    pages = "034",
    year = "2024"
}

@article{ATLAS:2025lfx,
    author = "Aad, Georges and others",
    collaboration = "ATLAS",
    title = "{Search for emerging jets in $pp$ collisions at $\sqrt{s} = 13$ TeV with the ATLAS experiment}",
    eprint = "2510.12347",
    archivePrefix = "arXiv",
    primaryClass = "hep-ex",
    reportNumber = "CERN-EP-2025-225",
    doi = "10.1140/epjc/s10052-026-15693-1",
    journal = "Eur. Phys. J. C",
    volume = "86",
    pages = "808",
    year = "2026"
}

@article{CMS:2024gxp,
    author = "Hayrapetyan, Aram and others",
    collaboration = "CMS",
    title = "{Search for dark QCD with emerging jets in proton-proton collisions at $ \sqrt{s} $ = 13 TeV}",
    eprint = "2403.01556",
    archivePrefix = "arXiv",
    primaryClass = "hep-ex",
    reportNumber = "CMS-EXO-22-015, CERN-EP-2024-049",
    doi = "10.1007/JHEP07(2024)142",
    journal = "JHEP",
    volume = "07",
    pages = "142",
    year = "2024"
}

@techreport{CMS-PAS-EXO-24-010,
      collaboration = "CMS",
      title         = "{Search for nonresonant production of strongly coupled
                       dark matter in proton-proton collisions at
                       $\sqrt{s}=13~\mathrm{TeV}$}",
      institution   = "CERN",
      reportNumber  = "CMS-PAS-EXO-24-010",
      address       = "Geneva",
      year          = "2026",
      url           = "https://cds.cern.ch/record/2967685",
}

\end{document}